\RequirePackage[hyphens]{url}
\documentclass[journal=jpccck,manuscript=article]{achemso}

\usepackage[version=3]{mhchem} 
\usepackage{xcolor}
\usepackage{soul} 
\usepackage{amssymb}
\usepackage{lineno,hyperref}
\usepackage{graphicx}
\usepackage{dcolumn}
\usepackage{bm}
\usepackage{lipsum}

\newcommand{\DIPC}[0]{
Donostia International Physics Center (DIPC),
Paseo Manuel de Lardizabal 4, 20018 Donostia-San Sebasti\'an, Spain}
\newcommand{\CFM}[0]{
Centro de F\'{\i}sica de Materiales CFM/MPC (CSIC-UPV/EHU), Paseo Manuel de Lardizabal 5, 20018 Donostia-San Sebasti\'an, Spain}
\newcommand{\burdeos}[0]{Institut des Sciences Moléculaires (ISM), Université de Bordeaux, 351 Cours de la Libération, 33405 Talence, France}

\author{Raúl Bombín}
\email{raul.bombin@u-bordeaux.fr}
\affiliation{\burdeos}

\author{Ricardo Díez Muiño}
\email{rdm@ehu.eus}
\affiliation{\DIPC}

\author{J.\ Iñaki Juaristi}
\email{josebainaki.juaristi@ehu.eus}
\affiliation{Departamento de Pol\'{\i}meros y Materiales Avanzados: F\'{\i}sica, Qu\'{\i}mica y Tecnolog\'{\i}a, Facultad de Qu\'{\i}micas (UPV/EHU), Apartado 1072, 20080 Donostia-San Sebasti\'an, Spain}
\alsoaffiliation{\CFM}
\alsoaffiliation{\DIPC}

\author{Maite Alducin}
\email{maite.alducin@ehu.eus}
\affiliation{\CFM}
\alsoaffiliation{\DIPC}

\title{Scattering of CO from Vacant-MoSe$_2$ with O Adsorbates: Is CO$_2$ Formed?}

\date{\today}

\begin{document}

\begin{tocentry}

\includegraphics[width=1\columnwidth]{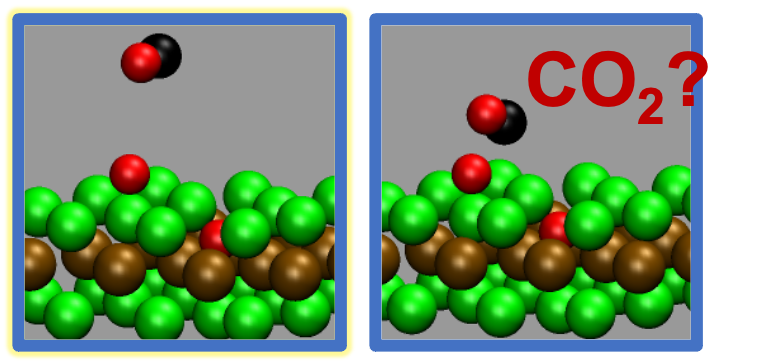}

\end{tocentry}

\begin{abstract}
Using ab initio molecular dynamics (AIMD) simulations, based on density functional theory that also accounts for van der Waals interactions, we study the oxidation of gas phase CO on MoSe$_2$ with a Se vacancy and oxygen coverage of 0.125~ML. In the equilibrium configuration, one of the O atoms is adsorbed on the vacancy and the other one atop one Se atom. Recombination of the CO molecule with the second of these O atoms to form CO$_2$ is a highly exothermic reaction, with an energy gain of around 3~eV. The likeliness of the CO oxidation reaction on this surface is next examined by calculating hundreds of AIMD trajectories for incidence energies that suffice to overcome the energy barriers in the entrance channel of the CO oxidative recombination. In spite of it, no CO$_2$ formation event is obtained. In most of the calculated trajectories the incoming CO molecule is directly reflected and in some cases, mainly at low energies, the molecules remain trapped at the surface but without reacting. As important conclusion, our AIMD simulations show that the recombination of CO molecules with adsorbed O atoms is a very unlikely reaction in this system, despite its large exothermicity.
\end{abstract}



\section{Introduction \label{sec:intro}}

In the last years, the possibility of using two-dimensional (2D) materials for catalysis~\cite{Singh2021,Giuffredi2021,Wazir2022,Adabala2022,Arumugam2024,Khaidar2024} and gas sensing~\cite{Tang2022,Tsai2022,Sulleiro2022,Ottaviano2023} has been broadly explored. 
Different families of 2D materials, like transition metal carbides or nitrides~\cite{SaiBhargava2024} and transition metal dichalcogenides (TMD)~\cite{Liu2017,Ping2017,Zeng2018,Bernal2019,Tyagi2020,Zheng2021,Zhai2024}, have been considered for these purposes. 
The motivation roots in the exceptional properties that these materials exhibit for such applications, as for example, a large surface to volume ratio and good electronic and optical properties.  
Importantly, their properties can be tuned by several procedures, including application of strain \cite{Muoi2019}, application of an external electric field~\cite{Ai2019}, creation of vacancies~\cite{Santos2020,Liang2021,Koos2019,Zhou2021HER,Lunardon2023}, stacking different two-dimensional materials~\cite{Feng2016,Ottaviano2023}, and even by considering non 2D configurations~\cite{YANG2020127369,Zhou2021flower,Jaiswal2022}. The latter permits to engineer new 2D materials with the desired properties.
For gas sensing and electrocatalysis applications, TMD semiconductors, in which we focus on in this work, offer the advantage of being easily implemented with electrodes\cite{Tyagi2020}, while overcoming the zero band-gap problem of graphene that complicates its implementation in electronic circuits~\cite{Guanxiong2013}.

Focusing on the usage of TMDs as catalysts for diverse chemical processes, their applicability is  promising. There are various studies exploring the catalytic properties of TMDs for the hydrogen evolution reaction (HER)~\cite{Shu2017,Lee2018,Jain2020,Zhou2021HER,Wazir2022,Yi2019,Lunardon2023} and the oxygen evolution reaction (OER)~\cite{German2020,Karmodak2021,Mohanty2018}. These are the two reactions involved in water splitting, a key process for sustainable energy production. 
In the case of HER, TMD materials appear as good alternatives to precious metals such as platinum. However, 
pristine TMDs exhibit poor performance for OER due to the large binding energies of intermediate products on the TMD surface~\cite{German2020}. 
Nonetheless, both theory and experiments reveal that the efficiency of these two reactions can be enhanced at the edges of the nanosheets~\cite{Mohanty2018,Karmodak2021,Wazir2022}, around chalcogen and metal vacancies~\cite{Shu2017,Lee2018,Zhou2021HER}, and by doping the nanosheet~\cite{Jain2020}. Other chemical reactions that are the focus on theoretical studies of the TMD catalytic properties 
include, for example,  CO$_2$ reduction  to produce acids and other valuable products~\cite{chan2014,AGUILAR2020147611,Kang2019COr,Ye2021,Ren2021,Ramzan2024,Singh2021,Giuffredi2021,Khaidar2024}, N$_2$ reduction to produce ammonia~\cite{Zhang2020}, production of ethanol from syngas~\cite{Qu2020}, and also CO oxidation on MoS$_2$ doped with metal atoms~\cite{MA2015,Zhu2019}, on PtTe$_2$ in the presence of O$_2$~\cite{Zhang2020COox}, and on the Janus WSSe surface~\cite{Guo2021}. 

In the aforereviewed bibliography, the catalytic properties are commonly determined in terms of the adsorption energies and minimum energy paths of the reaction considered. Being the latter the common procedure, the purpose and novelty of our work are to explore the catalytic properties of MoSe$_2$ for CO$_2$ formation but from a different perspective, the reaction dynamics. DFT calculations including van der Waals corrections showed that the dissociative adsorption of O$_2$, which is endothermic by more than 1~eV in the pristine 1H-MoSe$_2$ sheet, becomes highly exothermic in the presence of Se vacancies~\cite{bombin2022}. In light of those results, here we focus on the dynamics of the recombination process between one O atom adsorbed on the MoSe$_2$ surface with a Se vacancy and a gas-phase CO molecule. This kind of reaction process, which can occur upon a single collision of the gas species with either the adsorbate (direct Eley-Rideal) or the surface (indirect Eley-Rideal) or after few collisions that serve to slow down the incoming atom/molecule (hot atom/molecule abstraction), are expected to efficiently compete with the thermal Langmuir-Hinshelwood recombination when the adsorbate is strongly bound to the surface. This is precisely the case in the dissociative adsorption of O$_2$ on the MoSe$_2$ surface with a Se vacancy. As shown in ref~\citenum{bombin2022}, the adsorption energies of the nascent O atoms formed upon the dissociative adsorption of O$_2$ in that surface are as strong as $-2.9$~eV and $-7.3$~eV, depending on the adsorption site. Our new calculations for the recombinative formation of CO$_2$ via an Eley-Rideal mechanism show that the reaction is energetically favored and thus, motivated us to study the process with ab initio molecular dynamics simulations (AIMD). Interestingly, none of the incoming CO molecules is able to abstract an O adatom and form CO$_2$, even if the molecule has enough energy to overcome the entrance barriers for the reaction. These results remark the importance of going beyond static calculations of the reaction energetics in determining the catalyzing character of the material under consideration.

The paper is organized as follows, the general computational settings employed to both model the system and simulate the reaction dynamics with the AIMD methodology are detailed in the Theoretical Methods section. The Results \& Discussion section starts with the analysis of the energetics of the main reactions that may occur in the system for the incidence conditions considered, i.e., a gas-phase CO impinging on the MoSe$_2$ surface with a Se vacancy and an oxygen coverage of 0.125~ML. Next, the AIMD results are presented and discussed in light of the information extracted from the in depth analysis of the trajectories. The summary and conclusions of our work are given in the last section.

\section{Theoretical Methods \label{sec:methods}}

\subsection{General DFT Computational Settings}

All DFT calculations are performed with the Vienna \textit{ab initio} simulation package ({\sc vasp}) \cite{Kresse1996} using the revPBE-vdW exchange-correlation functional proposed by Dion {\it et al}.~\cite{Dion2004}. The ionic cores are described with the projector augmented wave (PAW) method \cite{blockl1994} implemented in VASP~\cite{Kresse1999}. Specifically, we use the PAW potentials for Mo, Se, O, and C that have 14, 6, 6, and 4 valence electrons, respectively. The Kohn-Sham states are expanded in a plane-wave basis set using an energy cutoff of 700~eV. The Brillouin zone integration is performed with a $\Gamma$-centered 4$\times$4$\times$1 Monkhorst–Pack grid of special $\mathbf{k}$-points~\cite{Monkhorst1976}. A Gaussian smearing of 0.1~eV is used for electronic state occupancies. 

Calculations are performed for the $1H$ phase of the MoSe$_2$ monolayer with a small concentration of non-interacting Se vacancies (denoted as MoSe$_2$-vSe hereafter). Following ref~\citenum{bombin2022}, the latter is adequately described by a 4$\times$4 surface supercell that contains a single Se vacancy and is separated from its periodic image by 23~{\AA} of vacuum along the surface normal. The value of the hexagonal lattice constant is $a$=3.386~{\AA}~\cite{bombin2022}. The O$_2$ molecule is dissociatively adsorbed in the configuration analyzed in ref~\citenum{bombin2022}. Namely, one of the O atoms is adsorbed on top of the Se vacancy and the other one on top of the closest Se atom. The O atom in the vacancy is located at $z=-0.62$~{\AA} below the surface, the O atom on top of the Se atom is located at $z$=1.57~{\AA} above the surface, and the distance between them is d(O-O)=4.01~{\AA}. Dissociation of the O$_2$ molecule in this configuration is exothermic by 3.3~eV. We denote this structure as 2O-MoSe$_2$-vSe.

Regarding our choice of the revPBE-vdW functional, our previous study of the adsorption properties of different diatomic molecules on MoSe$_2$ and MoSe$_2$-vSe showed that both revPBE-vdW and PBE-D3~\cite{Grime2010} provide a satisfactory description of the experimental MoSe$_2$ geometrical and electronic structures. Furthermore, even if PBE-D3 tends to overbinding as compared to revPBE-vdW, there were no significant qualitative differences between both functionals in the O, O$_2$, and CO adsorption and dissociation energies and adsorption sites on MoSe$_2$ and MoSe$_2$-vSe.

\subsection{AIMD Initial Conditions}

The interaction dynamics of CO with the 2O-MoSe$_2$-vSe surface is studied with total energy-conserving AIMD simulations for an initial substrate temperature of 100~K. The initial positions and velocities of the atoms in the substrate are randomly taken from a canonical ensemble of configurations. The latter is generated by thermalizing the 2O-MoSe$_2$-vSe substrate at 100~K during more than 10~ps using the Nos\'e-Hoover thermostat~\cite{Nose84} implemented in VASP. The use of constant-energy AIMD simulations with an appropriate set of initial positions and velocities constitutes a reasonable approximation to incorporate surface temperature effects during the short simulation time of interest (up to 1.0--1.4~ps in our case)~\cite{Novko2017}. This is a common procedure in gas-surface dynamics simulations~\cite{Gross07, Nattino12, Nattino14,Kolb16,Novko2017,Zhou17,Zhoujcp18b,Fuchseljpcc19,santamaria2019,santamaria2021,GRANJADELRIO2021}.

The CO molecule starts with its center of mass (CM) located at a distance $Z_\textrm{cm}=6$~{\AA} from the 2O-MoSe$_2$-vSe surface with the vibrational zero-point energy and zero rotational energy. 
The initial molecular orientation and CM position in the $(x,y)$-plane parallel to the surface are randomly sampled. The ($X_\textrm{cm}$,$Y_\textrm{cm}$)-sampling however focuses on the simulation cell area that contains the O adatoms with the purpose of optimizing the computational effort (see below). The molecule impinges perpendicular to the surface (i.e., normal incidence). Simulations are performed for five different initial translational energies, namely, $E_{i}$=0.2, 0.4, 0.6, 0.8, and 1.0~eV. For each $E_{i}$, 200 AIMD trajectories are run using the Beeman predictor-corrector algorithm with an integration time step of 1~fs. At the end of each AIMD trajectory, the impinging CO is considered reflected when its center of mass reaches a distance from the Se topmost layer $Z_\textrm{cm}\geq 6$~{\AA} with positive velocity along the surface normal ($P_\textrm{cm,z}>0$). It is considered trapped if after reaching the maximum integration time of 1~ps, the previous condition was not met. In the case of recombinative CO$_2$ formation, the same criteria is used to  distinguish between CO$_2$ desorption and trapping. 

\section{Results \& Discussion\label{sec:results}}

\begin{figure}
\includegraphics*[angle=0,width=0.9\columnwidth]{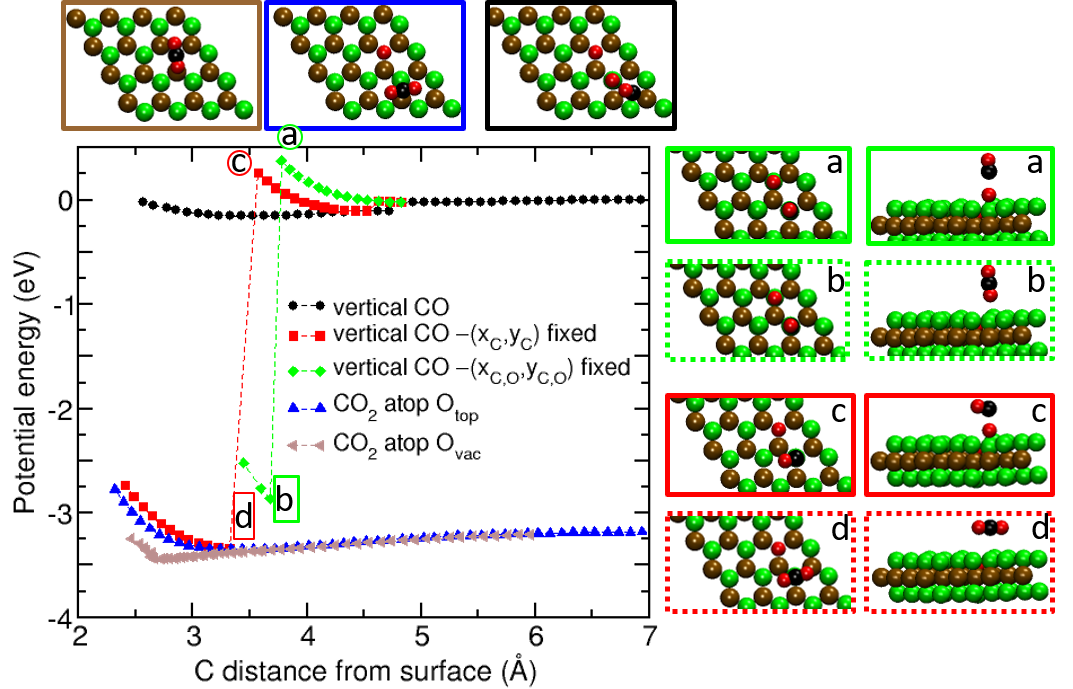}%
\caption{Potential energy against the C distance from the topmost Se layer $z_\textrm{C}$. The zero reference energy corresponds to CO in the middle of vacuum. Black, red, and green curves correspond to a vertically oriented CO approaching the surface above O$_\textrm{top}$ under different relaxation conditions (see text for details). Brown and blue curves correspond, respectively, to the CO+O$_\textrm{top}$-recombined CO$_2$ approaching the surface over O$_\textrm{vac}$ and over the Se atom on which O$_\textrm{top}$ was adsorbed. The three structures on the top and from left to right correspond to the minimum energy configurations for the curves labeled CO$_2$ atop O$_\textrm{vac}$ (brown curve), CO$_2$ atop O$_\textrm{top}$ (blue curve), and vertical CO (black curve). Top and side views of the structures before and after the discontinuities in the red (points labeled c and d) and green (points labeled a and b) vertical CO curves are shown in the right panels following the same labeling. Se (top layer), Mo, O, and C are plotted in green, brown, red, and black, respectively. 
}
\label{fig:1dpes}
\end{figure}

The calculated reaction energies summarized in table~\ref{tab:reaction_energies} show that both the recombinative CO$_2$ adsorption ($-3.437$~eV) and abstraction ($-3.174$~eV) involving the O atom adsorbed atop Se (O$_\textrm{top}$ hereafter) are highly exothermic. Recombination with the O atom adsorbed in the Se vacancy (O$_\textrm{vac}$ hereafter) is energetically less favored. In particular, adsorption is exothermic by only $-0.286$~eV, while abstraction becomes endothermic, being the reaction energy $1.297$~eV. 
The different adsorption energy for O$_\textrm{top}$ ($-2.872$~eV) and O$_\textrm{vac}$ ($-7.300$~eV)~\cite{bombin2022} is the main reason for the different reaction energies obtained with each adsorbate. CO adsorption near O$_\textrm{top}$ although exothermic is energetically less favored than CO$_2$ formation. In this adsorption configuration (denoted as CO$_\textrm{h}$ in table~\ref{tab:reaction_energies}), the CO center of mass is practically over a hollow site (see the black-framed panel in Figure~\ref{fig:1dpes}). Interestingly, the calculated CO$_\textrm{h}$ adsorption energy on 2O-MoSe$_2$-vSe ($-0.154$~eV) is similar to that of CO adsorbed over a hollow site on pristine MoSe$_2$ ($-0.143$~eV)~\cite{bombin2022}. Nonetheless, since the focus of this work is the recombinative formation of CO$_2$, we did not perform a systematic analysis of all possible CO adsorption configurations on 2O-MoSe$_2$-vSe and, therefore, we cannot discard the existence of CO adsorption configurations more stable than CO$_\textrm{h}$. 
\begin{table}
    \centering
    \begin{tabular}{c c c}
        \hline\hline
        Reaction  & Reactant/Product Adsorption Site         & $E$ (eV)\\
        \hline
        CO$_\textrm{(g)}$ + O$_\textrm{(a)} \rightharpoonup $ CO$_{2\textrm{(g)}}$          & O$_\textrm{top}$/--           & $-3.174$  \\ 
        CO$_\textrm{(g)}$ + O$_\textrm{(a)} \rightharpoonup $ CO$_{2\textrm{(g)}}$          & O$_\textrm{vac}$/--           & $1.297$  \\ 
        CO$_\textrm{(g)}$ + O$_\textrm{(a)} \rightharpoonup $ CO$_{2\textrm{(a)}}$          & O$_\textrm{top}$/ CO$_{2 \;\textrm{O}_\textrm{vac}}$          & $-3.437$  \\ 
        CO$_\textrm{(g)}$ + O$_\textrm{(a)} \rightharpoonup $ CO$_{2\textrm{(a)}}$          & O$_\textrm{vac}$/ CO$_{2 \;\textrm{O}_\textrm{vac}}$          & $-0.286$  \\ 
        CO$_\textrm{(g)}  \rightharpoonup $ CO$_\textrm{(a)}$          &  --/CO$_\textrm{h}$           & $-0.154$  \\

    \hline\hline
    \end{tabular}
   \caption{Reaction energies $E$ for CO$_\textrm{(g)}$+O$_\textrm{(a)}$ recombinative abstraction, CO$_\textrm{(g)}$+O$_\textrm{(a)}$ recombinative adsorption, and  CO$_\textrm{(g)}$ adsorption on the 2O-MoSe$_2$-vSe surface. Corresponding reactant/product adsorption sites are also indicated. The subscripts (g) and (a) stand for ``gas" and ``adsorbed", respectively.}
    \label{tab:reaction_energies}
\end{table}

In addition to calculating the reaction energy, 
we carried out a preliminary inspection of the possible energy barriers the incoming CO may encounter while approaching the surface. For simplicity, we focused on exploring a few paths running close to O$_\textrm{top}$. In these calculations the height of the C atom $z_\textrm{C}$ (and the 3 coordinates of one Mo atom located more than 7~{\AA} away the O adsorbates) are always fixed. In some paths, also the $(x,y)$ coordinates of the C and O atoms forming the molecule are fixed. The rest of degrees of freedom are allowed to relax until forces are smaller than 0.015~eV/{\AA}. The dependence of the potential energy on $z_\textrm{C}$ along these inspection paths is represented in Figure~\ref{fig:1dpes}. A vertically oriented CO approaching atop O$_\textrm{top}$ will face an energy barrier of about 0.378~eV, if the CO molecule is forced to stay perpendicular to O$_\textrm{top}$ by fixing the center of mass coordinates $(X_\textrm{cm}, Y_\textrm{cm})$ (green diamonds). The abrupt energy decrease at $z_\textrm{C}\simeq 3.5$~{\AA} corresponds to abstraction of the adsorbed O to form  a vertically oriented CO$_2$ (see top and side views of the structures at points a and b in Figure~\ref{fig:1dpes}). Note that the energy at b is greater than that of desorbed CO$_2$ (asymptotic limit of the CO$_2$ curves plotted by blue and brown triangles in the figure), suggesting that an eventually formed vertical-CO$_2$ will end desorbing. The latter was confirmed with an ulterior structural relaxation of the system configuration at b, which ends with the CO$_2$ molecule in vacuum. A similar energy barrier ($0.260$~eV) is found when only the C coordinates ($x_\textrm{C}$, $y_\textrm{C}$, $z_\textrm{C}$) are fixed (red squares). In this case CO can tilt while approaching O$_\textrm{top}$ (see structure c in Figure~\ref{fig:1dpes}) and also ends forming CO$_2$ (see structure d), giving rise to the energy drop observed between points c and d in the figure. In this case the formed molecule is oriented parallel to the surface and it is energetically more stable than the desorbed CO$_2$. Finally, when only $z_\textrm{C}$ is fixed while approaching over O$_\textrm{top}$, CO tilts and is progressively repealed away the initial O$_\textrm{top}$--C vertical line, avoiding the energy barriers found in the two previous paths. The minimum energy configuration along this path is shown in the black framed panel. For completeness, we also calculate potential energy curves against $z_\textrm{C}$ for the CO$_2$ formed from CO$_\textrm{(g)}$+O$_\textrm{top}$. In these calculations only $z_\textrm{C}$ is fixed. Energies for CO$_2$ approaching over O$_\textrm{vac}$ and atop the Se atom on which O$_\textrm{top}$ was adsorbed prior recombining are respectively plotted by brown and blue curves in Figure~\ref{fig:1dpes}. Corresponding minimum energy structures are depicted in the brown and blue framed panels. 

Although CO$_2$ formation and desorption upon O$_\textrm{top}$ abstraction is energetically favored, the previous analysis casts some doubts regarding the likeliness of the reaction. The explored paths present energy barriers in the entrance channel that may difficult the required approach of the CO molecule to O$_\textrm{top}$ prior recombining. Nevertheless, we cannot exclude the existence of other complex but more favorable reaction paths with lower or nonexistent energy barriers. In this respect, there are many examples showing that the incoming gaseous species can also capture the adsorbate upon reflecting once (indirect ER recombination) or several times (hot atom/molecule recombination) with the surface. Both direct and indirect ER recombination are observed in molecular dynamics (MD) and AIMD simulations of H$_\textrm{(g)}$ scattered off Cl/Au(111)~\cite{Quattrucci2005, Zhou2018} and H/Cu(111)~\cite{Chen2019, Zhu2023} and in MD simulations of N$_\textrm{(g)}$ scattered off N-covered Ag(111)~\cite{Blanco-Rey2013, Juaristi2015} and N-covered W(110)~\cite{Quintas2012JCP, Petuya2014JPCC,Galparsoro2017}, whereas only indirect ER occurs in the  N-covered W(100)~\cite{Quintas2012JCP, Petuya2014JPCC} even at grazing incidence conditions at which direct ER is likely to occur~\cite{Galparsoro2017} and, important for the present study, in the formation of CO$_2$ by O$_\textrm{(g)}$ atoms impinging on CO/Pt(111)~\cite{Wu2019}. In the case of H$_\textrm{(g)}$ impinging on H-covered W(110) and W(100) surfaces, all ER recombinations are indirect~\cite{Galparsoro2017}, being also important the hot atom mechanism as shown in ulterior simulations performed at finite coverage~\cite{Galparsoro2016, GalparsoroJCPcom2017, Galparsoro2018}. 

\begin{figure}
\includegraphics*[angle=0,width=0.9\columnwidth]{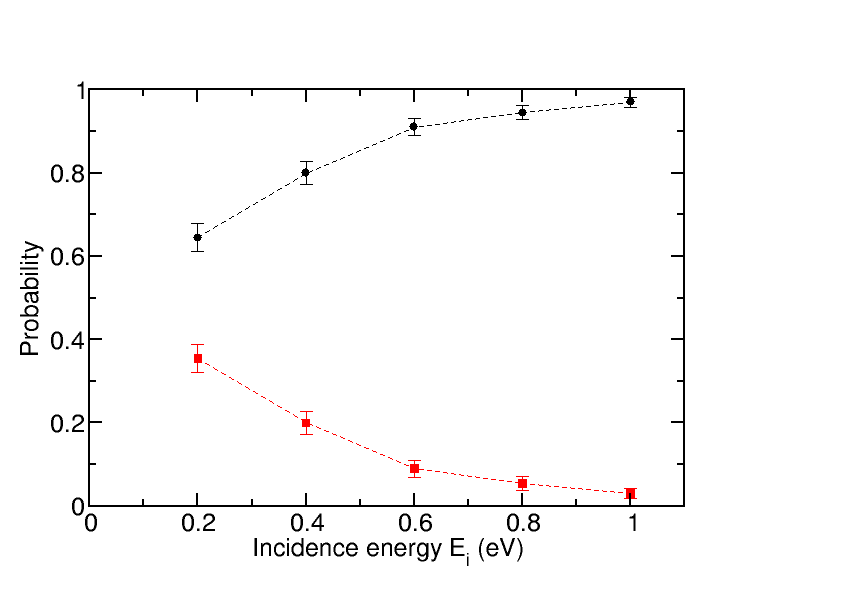}%
\caption{Scattering (black circles) and trapping (red squares) probability against incidence energy $E_i$ for CO impinging normal to the 2O-MoSe$_2$-vSe surface.
}
\label{fig:probability}
\end{figure}
Considering the rich reactivity observed in other systems, the results of our AIMD calculations are quite surprising. Even if most of the simulated incidence energies ($E_i=0.2-1.0$~eV) suffice to overcome the energy barriers explored in Figure~\ref{fig:1dpes}, none of the trajectories led to CO$_2$ formation. The latter includes both desorption and trapping (adsorption) of the expected CO$_\textrm{(g)}$+O$_\textrm{(a)}$ recombination. There are only two possible outcomes  observed at the end of simulations that correspond to the impinging CO being either reflected or molecularly trapped on the surface. Processes such as O$_\textrm{(a)}$ atomic desorption or O$_2$ recombinative desorption are highly endothermic and, therefore, energetically forbidden for the incidence energies here considered. Figure~\ref{fig:probability} shows the probabilities for CO reflection and trapping as a function of the incidence energy $E_i$. At the lowest considered energy ($E_i=0.2$~eV) 65\% of the molecules are reflected, whereas a high 35\% of them remain dynamically trapped at the end of the simulations (1~ps). Once the first collision with the surface takes place within a short interval of 200--400~fs, the trapped CO molecules experience a second rebound in the vacuum side at very diverse heights ($3.5 \leq Z_\textrm{cm} < 6$~{\AA}) and are attracted back to the surface. After extending the integration time up to about 1.4~ps, all these molecules stay below 6~{\AA} from the surface, but the final outcome of these trajectories at longer times remains unclear with our limited-time dynamics (see Figure S1 in the Supporting Information). As $E_i$ increases, the trapping probability rapidly decreases in favor of CO reflection. Thus, at the largest $E_i=1$~eV more than 95\% of the molecules are reflected. With very few exceptions, observed at lower incidence energies, the reflection process occurs after the very first collision with the surface. For $E_i=0.2, 0.4$~eV, the first (classical) turning point of the reflected molecules is distributed in three regions depending on whether CO collides near O$_\textrm{top}$ ($Z_\textrm{cm} \geq 3.5$~{\AA} ), near O$_\textrm{vac}$ ($Z_\textrm{cm} \leq 2.25$~{\AA} ), or at other points on the surface ($ 2.25 < Z_\textrm{cm} < 3.5$~{\AA}). These turning point regions move closer to the surface and broaden as $E_i$ increases. The turning point distributions for the trapped molecules exhibit a less featured structure. All these observations can be seen in Figures S2, S3, and S4 in the Supporting Information. 

\begin{figure}
\includegraphics*[angle=0,width=1\columnwidth]{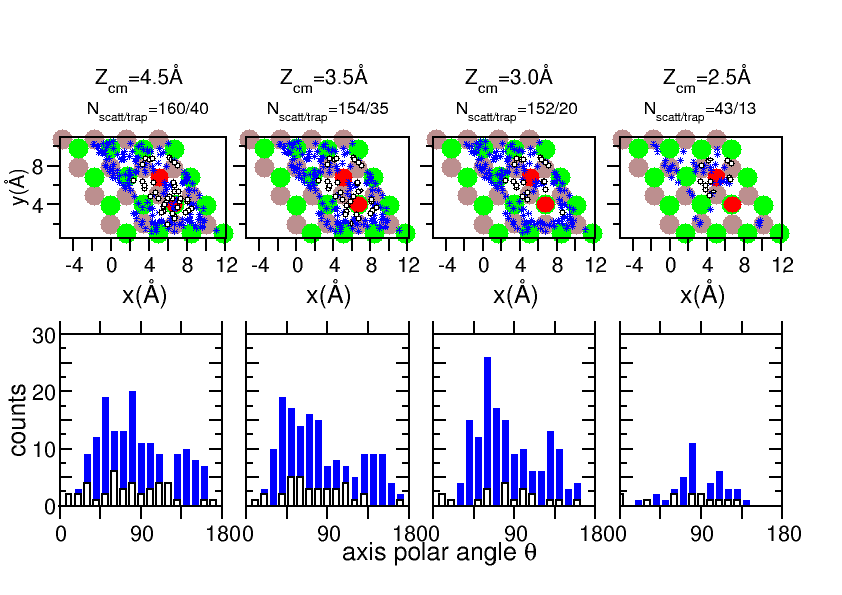}%
\caption{Evolution of the reflected (blue crosses and bars) and trapped (black empty circles and bars) CO molecules along the incoming trajectory impinging normal to 2O-MoSe$_2$-vSe with an initial incidence kinetic energy $E_i=0.4$~eV. Top: CO center of mass position over the surface unit cell when first reaching the height $Z_\textrm{cm}$ given above each column together with the number of reflected and trapped CO that reach that height. Bottom: polar angle distribution of CO molecular axis for a 10$^\circ$ binning. Se (top layer) and Mo surface atoms and O adatoms are represented by green, brown, and red circles, respectively.
}
\label{fig:xy_0.4eV}
\end{figure}
%
\begin{figure}
\includegraphics*[angle=0,width=1\columnwidth]{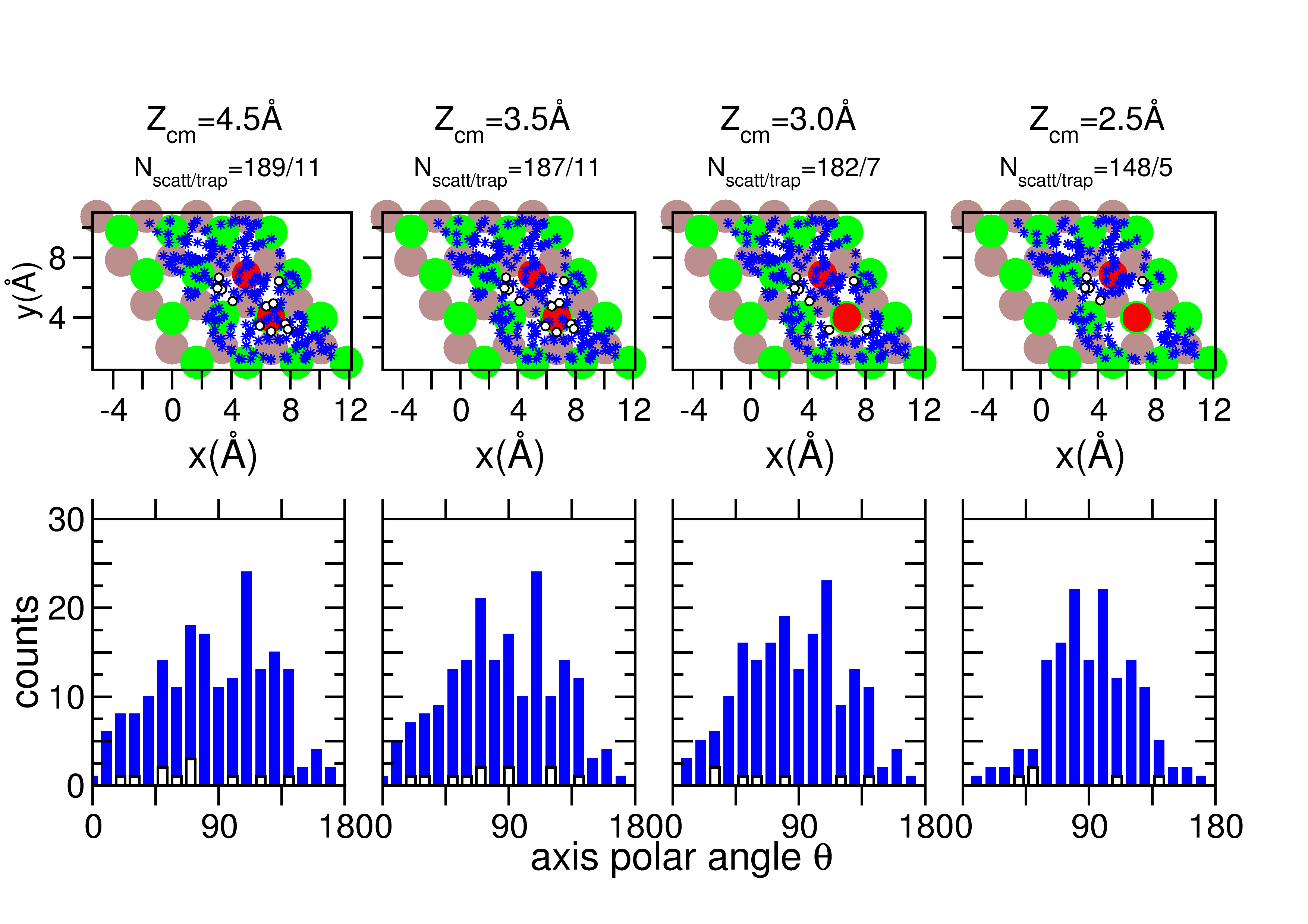}%
\caption{Same as Figure~\ref{fig:xy_0.4eV} but for initial incidence kinetic energy $E_i=0.8$~eV.
}
\label{fig:xy_0.8eV}
\end{figure}
A more thorough analysis of the trajectories allows us to better understand the lack of recombination processes. Figures~\ref{fig:xy_0.4eV} and~\ref{fig:xy_0.8eV} show the evolution along the incoming trajectory of the CO center of mass coordinates ($X_\textrm{cm}, Y_\textrm{cm}$) and the polar angle w.r.t. the surface normal of the CO molecular axis $\theta$ for incidence energies $E_i=0.4$ eV and $E_i=0.8$~eV, respectively. (Note in passing that the center of mass position at $Z_\textrm{cm}=4.5$~{\AA} provides a clear idea of the area covered by the random distribution over the surface that was used in the initial conditions sampling described in sec AIMD Initial Conditions.) The number of reflected and trapped molecules that reach a certain height $Z_\textrm{cm}$ during the initial approach is written on each column, providing information on  how many molecules have experienced the first collision with the surface between two consecutive height values. The most salient feature in both figures is the inability of the CO molecules to get close enough to O$_\textrm{top}$. In the case of $E_i=0.4$ eV, already at $Z_\textrm{cm}=3.5$~{\AA}, molecules that were impinging nearby O$_\textrm{top}$ have been reflected (6 molecules) or have been deviated to remain trapped at the surface (5 molecules). At the higher energy $E_i=0.8$~eV (Figure~\ref{fig:xy_0.8eV}), the molecules can get closer. Yet two of the molecules incoming above O$_\textrm{top}$ have already been reflected at $Z_\textrm{cm}=$~3.5~{\AA} and at $Z_\textrm{cm}=$~3~{\AA} all the molecules that were approaching O$_\textrm{top}$ have been reflected or deviated from the  O$_\textrm{top}$ proximity in the case of the trapped molecules. Interestingly, a majority of the molecules that can get as close to the surface as $Z_\textrm{cm}=2.5$~{\AA} are the ones that are in the region close to O$_\textrm{vac}$. However, O$_\textrm{vac}$ is buried in the Se topmost layer ($-0.62$~{\AA})~\cite{bombin2022} and strongly bound to the substrate, such that recombination does not take place. Regarding the $\theta$ distributions, there is no noticeable effect related to the orientation of the molecules, except for a certain predominance of near parallel orientations ($45^{\circ} <\theta < 135^{\circ}$) at the closest distances. 

These results are consistent with the information obtained from the reaction paths analyzed in Figure~\ref{fig:1dpes}. The reaction is highly exothermic and direct once the CO molecule is close to O$_\textrm{top}$, with the C atom at a distance from the surface below 3.5--4 \AA~depending on its orientation. However, the energy barriers that exist in the incoming path prevent the molecules from reaching that configuration. As a consequence, the CO molecules either reflect directly on top of O$_\textrm{top}$ before getting close enough to react, or are deviated and remain trapped at the surface. Though we cannot completely exclude that the trapped molecules may eventually get close to the O$_\textrm{top}$ in a proper orientation and react, the probability for this process is expected to be very low, considering that it has not occurred in any of the trapped trajectories for which the integration time has been increased up to 1.2--1.4~ps. In any case, in spite of its high exothermicity, direct ER recombination between CO impinging from vacuum and O$_\textrm{top}$ can apparently be excluded. To further support this conclusion, one additional trajectory was run for each $E_i$, in which the initial CO coordinates correspond to a vertically oriented CO located exactly above O$_\textrm{top}$ at height $Z_\textrm{cm}\simeq 4.5$~{\AA}. The turning point (taken as the height of the C atom) decreases to 2.49~{\AA} for $E_i=0.2$~eV and to 2.03~{\AA} for $E_i=1.0$~eV but in all cases the CO molecule bounces back into vacuum highly rotationally excited without forming a stable O$_\textrm{top}$--C bound.

Regarding the difference between reflected and trapped CO, Figures~\ref{fig:xy_0.4eV} and \ref{fig:xy_0.8eV} suggest that those molecules initially impinging over or nearby O$_\textrm{top}$ are more likely to be trapped after the first collision with the surface. But there is also a certain amount of trapped molecules that were initially located nearby O$_\textrm{vac}$. The evolution of the molecular orientation features no apparent differences between reflected and trapped CO. Altogether the difference between direct reflection and trapping seems related to minor details of the potential energy felt by the incoming CO molecule.

\begin{figure}
\includegraphics*[angle=0,width=1.0\columnwidth]{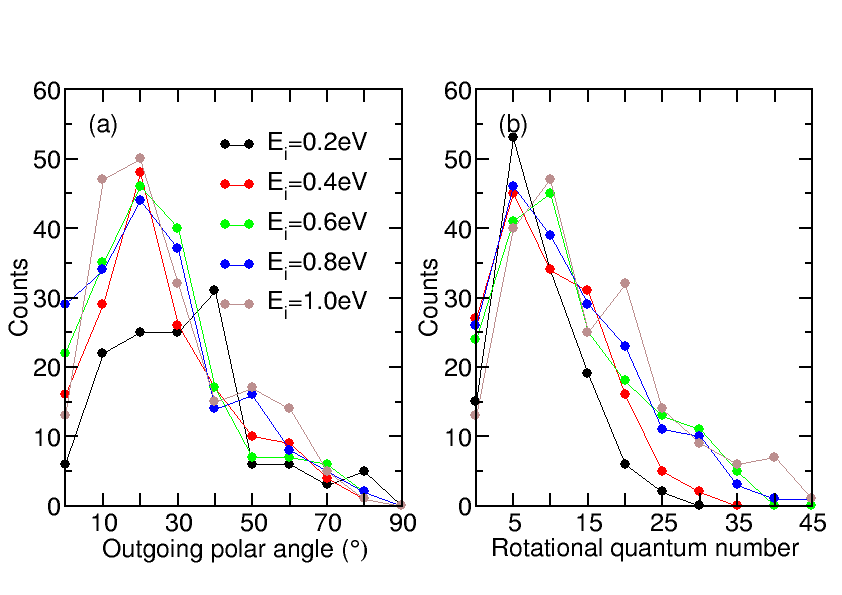}%
\caption{Outgoing polar angle (a) and rotational quantum number (b) distributions of the reflected CO molecules. Different colors correspond to different incidence energies $E_i$.
}
\label{fig:reflected}
\end{figure}
Finally, focusing on the reflected trajectories, in the left panel of Figure~\ref{fig:reflected} we observe that these molecules are predominantly reflected with small outgoing angles ($< 40^{\circ}$). This is expected for molecules under normal incidence that are directly reflected, i.e.,  without experiencing any relevant deviation along the incoming trajectory. The latter suggests a negligible corrugation of the potential energy surface as compared to the incidence energy $E_i$. In this respect, the wider distribution of the outgoing angle at $E_i=0.2$~eV agrees with the early energy barriers discussed above. 
Upon colliding with the surface, the reflected molecules become quite rotationally excited, particularly for incidence energies $E_i=0.6, 0.8$, and 1.0~eV. As shown in the right panel of Figure~\ref{fig:reflected}, the rotational quantum number reaches values as high as $J\geq 35$ for the largest studied $E_i$.

\section{Conclusions \label{sec:conclusions}}

We have studied the interaction of CO molecules coming from vacuum and a 2O-MoSe$_2$-vSe surface, in which one O atom is adsorbed in a Se vacancy and a second O atom is adsorbed on top of a Se atom of the $1H$ phase of the MoSe$_2$ monolayer. The latter is the equilibrium structure upon (dissociative) adsorption of a O$_2$ molecule on a MoSe$_2$ monolayer with one Se vacancy~\cite{bombin2022}. Interestingly, DFT calculations show that the CO recombination with the O atom adsorbed on top of the Se atom with either subsequent adsorption or desorption of the formed CO$_2$ molecules are highly exothermic reactions with energy gains of around 3--3.5 eV. Motivated by this finding, we have performed AIMD simulations with five different initial translational energies of CO molecules scattered off a 2O-MoSe$_2$-vSe surface thermalized at 100 K. Our aim was to analyze how likely was the CO oxidation in this system and the corresponding reaction mechanisms. However, none of the 200 AIMD trajectories run for each energy has led to CO$_2$ formation. The majority of the CO molecules are directly reflected, albeit in some cases (mainly at the lowest analyzed energies) some of them remain trapped at the surface without reacting with the adsorbed O atoms. The analysis of the trajectories show that those CO molecules impinging on  O$_\textrm{top}$ are reflected or deviated to trapping before they can get close enough to the O atom in order recombination can occur. Inspection of the possible reaction paths suggests that the configurational space of the oxidation process via an Eley-Rideal process is very restrictive. Due to the unavoidable limited statistics and integration time, which does not allow to completely determine the final outcome of the trapped CO molecules, we can not absolutely exclude the possibility of the reaction to occur at a later stage. Nevertheless, we can safely conclude that, in spite of its large exothermicity, the recombinative CO$_\textrm{(g)}$+O$_\textrm{(a)}$ abstraction is in the best case a very unlikely reaction in this system. As a general and important conclusion, the results of our dynamics simulations remark the importance of going beyond a pure static analysis of the reaction energetics to determine the catalytic character of any material.

\begin{acknowledgement}
Financial support  provided by the
Spanish MCIN$/$AEI$/$10.13039$/$501100011033$/$, FEDER Una manera de hacer Europa (Grant No.~PID2022-140163NB-I00), Gobierno Vasco-UPV/EHU (Project No.~IT1569-22), and the Basque Government Education Departments’ IKUR program, also co-funded by the European NextGenerationEU action through the Spanish Plan de Recuperación, Transformación y Resiliencia (PRTR). We also acknowledge the HPC resources provided by the Donostia International Physics Center (DIPC) Supercomputing Center. R.B. acknowledges funding from ADAGIO (Advanced Manufacturing Research Fellowship Programme in the Basque – New Aquitaine Region) from the  European Union’s Horizon 2020 research and innovation programme under the Marie Sklodowska  Curie cofund Grant Agreement No. 101034379.
\end{acknowledgement}

\begin{suppinfo}
The Supporting Information is available free of charge at XXX and includes detailed information of the turning point for the reflected and trapped molecules.

\end{suppinfo}


\bibliography{biblio2D}

\providecommand{\latin}[1]{#1}
\makeatletter
\providecommand{\doi}
  {\begingroup\let\do\@makeother\dospecials
  \catcode`\{=1 \catcode`\}=2 \doi@aux}
\providecommand{\doi@aux}[1]{\endgroup\texttt{#1}}
\makeatother
\providecommand*\mcitethebibliography{\thebibliography}
\csname @ifundefined\endcsname{endmcitethebibliography}
  {\let\endmcitethebibliography\endthebibliography}{}
\begin{mcitethebibliography}{82}
\providecommand*\natexlab[1]{#1}
\providecommand*\mciteSetBstSublistMode[1]{}
\providecommand*\mciteSetBstMaxWidthForm[2]{}
\providecommand*\mciteBstWouldAddEndPuncttrue
  {\def\EndOfBibitem{\unskip.}}
\providecommand*\mciteBstWouldAddEndPunctfalse
  {\let\EndOfBibitem\relax}
\providecommand*\mciteSetBstMidEndSepPunct[3]{}
\providecommand*\mciteSetBstSublistLabelBeginEnd[3]{}
\providecommand*\EndOfBibitem{}
\mciteSetBstSublistMode{f}
\mciteSetBstMaxWidthForm{subitem}{(\alph{mcitesubitemcount})}
\mciteSetBstSublistLabelBeginEnd
  {\mcitemaxwidthsubitemform\space}
  {\relax}
  {\relax}

\bibitem[Singh \latin{et~al.}(2021)Singh, Modak, Pant, Sinhamahapatra, and
  Biswas]{Singh2021}
Singh,~S.; Modak,~A.; Pant,~K.~K.; Sinhamahapatra,~A.; Biswas,~P.
  MoS$_2$–Nanosheets-Based Catalysts for Photocatalytic {CO$_2$} Reduction: A
  Review. \emph{ACS Appl. Nano Mater.} \textbf{2021}, \emph{4},
  8644--8667\relax
\mciteBstWouldAddEndPuncttrue
\mciteSetBstMidEndSepPunct{\mcitedefaultmidpunct}
{\mcitedefaultendpunct}{\mcitedefaultseppunct}\relax
\EndOfBibitem
\bibitem[Giuffredi \latin{et~al.}(2021)Giuffredi, Asset, Liu, Atanassov, and
  Di~Fonzo]{Giuffredi2021}
Giuffredi,~G.; Asset,~T.; Liu,~Y.; Atanassov,~P.; Di~Fonzo,~F. Transition Metal
  Chalcogenides as a Versatile and Tunable Platform for Catalytic {CO$_2$} and
  {N$_2$} Electroreduction. \emph{ACS Mater. Au} \textbf{2021}, \emph{1},
  6--36\relax
\mciteBstWouldAddEndPuncttrue
\mciteSetBstMidEndSepPunct{\mcitedefaultmidpunct}
{\mcitedefaultendpunct}{\mcitedefaultseppunct}\relax
\EndOfBibitem
\bibitem[Wazir \latin{et~al.}(2022)Wazir, Daud, Safeer, Almarzooqi, and
  Qurashi]{Wazir2022}
Wazir,~M.~B.; Daud,~M.; Safeer,~S.; Almarzooqi,~F.; Qurashi,~A. Review on 2D
  Molybdenum Diselenide (MoSe$_2$) and Its Hybrids for Green Hydrogen (H$_2$)
  Generation Applications. \emph{ACS Omega} \textbf{2022}, \emph{7},
  16856--16865\relax
\mciteBstWouldAddEndPuncttrue
\mciteSetBstMidEndSepPunct{\mcitedefaultmidpunct}
{\mcitedefaultendpunct}{\mcitedefaultseppunct}\relax
\EndOfBibitem
\bibitem[Adabala and Dutta(2022)Adabala, and Dutta]{Adabala2022}
Adabala,~S.; Dutta,~D.~P. A Review on Recent Advances in Metal
  Chalcogenide-Based Photocatalysts for {CO$_2$} Reduction. \emph{J. Environ.
  Chem. Eng.} \textbf{2022}, \emph{10}, 107763\relax
\mciteBstWouldAddEndPuncttrue
\mciteSetBstMidEndSepPunct{\mcitedefaultmidpunct}
{\mcitedefaultendpunct}{\mcitedefaultseppunct}\relax
\EndOfBibitem
\bibitem[Arumugam and Yang(2024)Arumugam, and Yang]{Arumugam2024}
Arumugam,~M.; Yang,~H.-H. A Review of the Application of Wide-Bandgap
  Semiconductor Photocatalysts for {CO$_2$} Reduction. \emph{J.{CO$_2$} Util.}
  \textbf{2024}, \emph{83}, 102808\relax
\mciteBstWouldAddEndPuncttrue
\mciteSetBstMidEndSepPunct{\mcitedefaultmidpunct}
{\mcitedefaultendpunct}{\mcitedefaultseppunct}\relax
\EndOfBibitem
\bibitem[Khaidar \latin{et~al.}(2024)Khaidar, Isahak, Ramli, and
  Ahmad]{Khaidar2024}
Khaidar,~D.~M.; Isahak,~W. N. R.~W.; Ramli,~Z. A.~C.; Ahmad,~K.~N. Transition
  Metal Dichalcogenides-Based Catalysts for {CO$_2$} Conversion: An Updated
  Review. \emph{Int. J. Hydrogen Energy} \textbf{2024}, \emph{68}, 35--50\relax
\mciteBstWouldAddEndPuncttrue
\mciteSetBstMidEndSepPunct{\mcitedefaultmidpunct}
{\mcitedefaultendpunct}{\mcitedefaultseppunct}\relax
\EndOfBibitem
\bibitem[Tang \latin{et~al.}(2022)Tang, Zhao, and Liu]{Tang2022}
Tang,~Y.; Zhao,~Y.; Liu,~H. Room-Temperature Semiconductor Gas Sensors:
  Challenges and Opportunities. \emph{ACS Sens.} \textbf{2022}, \emph{7},
  3582--3597, PMID: 36399520\relax
\mciteBstWouldAddEndPuncttrue
\mciteSetBstMidEndSepPunct{\mcitedefaultmidpunct}
{\mcitedefaultendpunct}{\mcitedefaultseppunct}\relax
\EndOfBibitem
\bibitem[Tsai \latin{et~al.}(2022)Tsai, Wang, Liu, Wang, and Huo]{Tsai2022}
Tsai,~H.-S.; Wang,~Y.; Liu,~C.; Wang,~T.; Huo,~M. The Elemental 2D Materials
  Beyond Graphene Potentially Used as Hazardous Gas Sensors for Environmental
  Protection. \emph{J. Hazard. Mater.} \textbf{2022}, \emph{423}, 127148\relax
\mciteBstWouldAddEndPuncttrue
\mciteSetBstMidEndSepPunct{\mcitedefaultmidpunct}
{\mcitedefaultendpunct}{\mcitedefaultseppunct}\relax
\EndOfBibitem
\bibitem[Sulleiro \latin{et~al.}(2022)Sulleiro, Dominguez-Alfaro, Alegret,
  Silvestri, and Gómez]{Sulleiro2022}
Sulleiro,~M.~V.; Dominguez-Alfaro,~A.; Alegret,~N.; Silvestri,~A.;
  Gómez,~I.~J. 2D Materials Towards Sensing Technology: From Fundamentals to
  Applications. \emph{Sens. Bio-Sens. Res.} \textbf{2022}, \emph{38},
  100540\relax
\mciteBstWouldAddEndPuncttrue
\mciteSetBstMidEndSepPunct{\mcitedefaultmidpunct}
{\mcitedefaultendpunct}{\mcitedefaultseppunct}\relax
\EndOfBibitem
\bibitem[Ottaviano and Mastrippolito(2023)Ottaviano, and
  Mastrippolito]{Ottaviano2023}
Ottaviano,~L.; Mastrippolito,~D. {The Future Ahead Gas Sensing with
  Two-Dimensional Materials}. \emph{Appl. Phys. Lett.} \textbf{2023},
  \emph{123}, 050502\relax
\mciteBstWouldAddEndPuncttrue
\mciteSetBstMidEndSepPunct{\mcitedefaultmidpunct}
{\mcitedefaultendpunct}{\mcitedefaultseppunct}\relax
\EndOfBibitem
\bibitem[{Sai Bhargava Reddy} and Aich(2024){Sai Bhargava Reddy}, and
  Aich]{SaiBhargava2024}
{Sai Bhargava Reddy},~M.; Aich,~S. Recent Progress in Surface and
  Heterointerface Engineering of 2D MXenes for Gas Sensing Applications.
  \emph{Coord. Chem. Rev.} \textbf{2024}, \emph{500}, 215542\relax
\mciteBstWouldAddEndPuncttrue
\mciteSetBstMidEndSepPunct{\mcitedefaultmidpunct}
{\mcitedefaultendpunct}{\mcitedefaultseppunct}\relax
\EndOfBibitem
\bibitem[Liu \latin{et~al.}(2017)Liu, Ma, Pinna, and Zhang]{Liu2017}
Liu,~X.; Ma,~T.; Pinna,~N.; Zhang,~J. Two-Dimensional Nanostructured Materials
  for Gas Sensing. \emph{Adv. Funct. Mater.} \textbf{2017}, \emph{27},
  1702168\relax
\mciteBstWouldAddEndPuncttrue
\mciteSetBstMidEndSepPunct{\mcitedefaultmidpunct}
{\mcitedefaultendpunct}{\mcitedefaultseppunct}\relax
\EndOfBibitem
\bibitem[Ping \latin{et~al.}(2017)Ping, Fan, Sindoro, Ying, and
  Zhang]{Ping2017}
Ping,~J.; Fan,~Z.; Sindoro,~M.; Ying,~Y.; Zhang,~H. Recent Advances in Sensing
  Applications of Two-Dimensional Transition Metal Dichalcogenide Nanosheets
  and Their Composites. \emph{Adv. Funct. Mater.} \textbf{2017}, \emph{27},
  1605817\relax
\mciteBstWouldAddEndPuncttrue
\mciteSetBstMidEndSepPunct{\mcitedefaultmidpunct}
{\mcitedefaultendpunct}{\mcitedefaultseppunct}\relax
\EndOfBibitem
\bibitem[Zeng \latin{et~al.}(2018)Zeng, Lin, Gu, and Li]{Zeng2018}
Zeng,~Y.; Lin,~S.; Gu,~D.; Li,~X. Two-Dimensional Nanomaterials for Gas Sensing
  Applications: The Role of Theoretical Calculations. \emph{Nanomaterials}
  \textbf{2018}, \emph{8}, 851\relax
\mciteBstWouldAddEndPuncttrue
\mciteSetBstMidEndSepPunct{\mcitedefaultmidpunct}
{\mcitedefaultendpunct}{\mcitedefaultseppunct}\relax
\EndOfBibitem
\bibitem[Vargas-Bernal(2019)]{Bernal2019}
Vargas-Bernal,~R. Electrical Properties of Two-Dimensional Materials Used in
  Gas Sensors. \emph{Sensors} \textbf{2019}, \emph{19}, 1295\relax
\mciteBstWouldAddEndPuncttrue
\mciteSetBstMidEndSepPunct{\mcitedefaultmidpunct}
{\mcitedefaultendpunct}{\mcitedefaultseppunct}\relax
\EndOfBibitem
\bibitem[Tyagi \latin{et~al.}(2020)Tyagi, Wang, Huang, Hu, Tang, Guo, Ouyang,
  and Zhang]{Tyagi2020}
Tyagi,~D.; Wang,~H.; Huang,~W.; Hu,~L.; Tang,~Y.; Guo,~Z.; Ouyang,~Z.;
  Zhang,~H. Recent Advances in Two-Dimensional-Material-Based Sensing
  Technology Toward Health and Environmental Monitoring Applications.
  \emph{Nanoscale} \textbf{2020}, \emph{12}, 3535--3559\relax
\mciteBstWouldAddEndPuncttrue
\mciteSetBstMidEndSepPunct{\mcitedefaultmidpunct}
{\mcitedefaultendpunct}{\mcitedefaultseppunct}\relax
\EndOfBibitem
\bibitem[Zheng \latin{et~al.}(2021)Zheng, Liu, Xie, Lu, and Zhang]{Zheng2021}
Zheng,~W.; Liu,~X.; Xie,~J.; Lu,~G.; Zhang,~J. Emerging van der Waals Junctions
  Based on TMDs Materials for Advanced Gas Sensors. \emph{Coordin. Chem. Rev.}
  \textbf{2021}, \emph{447}, 214151\relax
\mciteBstWouldAddEndPuncttrue
\mciteSetBstMidEndSepPunct{\mcitedefaultmidpunct}
{\mcitedefaultendpunct}{\mcitedefaultseppunct}\relax
\EndOfBibitem
\bibitem[Zhai \latin{et~al.}(2024)Zhai, Li, Wang, Zhai, Yao, Li, Wang, Yang,
  Chi, Liang, Shi, Ge, Lai, Yun, Zhang, Wu, He, Chen, Huang, and
  Zhang]{Zhai2024}
Zhai,~W.; Li,~Z.; Wang,~Y.; Zhai,~L.; Yao,~Y.; Li,~S.; Wang,~L.; Yang,~H.;
  Chi,~B.; Liang,~J. \latin{et~al.}  Phase Engineering of Nanomaterials:
  Transition Metal Dichalcogenides. \emph{Chem. Rev.} \textbf{2024},
  \emph{124}, 4479--4539, PMID: 38552165\relax
\mciteBstWouldAddEndPuncttrue
\mciteSetBstMidEndSepPunct{\mcitedefaultmidpunct}
{\mcitedefaultendpunct}{\mcitedefaultseppunct}\relax
\EndOfBibitem
\bibitem[Muoi \latin{et~al.}(2019)Muoi, Hieu, Phung, Phuc, Amin, Hoi, Hieu,
  Nhan, Nguyen, and Le]{Muoi2019}
Muoi,~D.; Hieu,~N.~N.; Phung,~H.~T.; Phuc,~H.~V.; Amin,~B.; Hoi,~B.~D.;
  Hieu,~N.~V.; Nhan,~L.~C.; Nguyen,~C.~V.; Le,~P. Electronic Properties of
  {WS$_2$} and {WSe$_2$} Monolayers with Biaxial Strain: A First-Principles
  Study. \emph{Chem. Phys.} \textbf{2019}, \emph{519}, 69--73\relax
\mciteBstWouldAddEndPuncttrue
\mciteSetBstMidEndSepPunct{\mcitedefaultmidpunct}
{\mcitedefaultendpunct}{\mcitedefaultseppunct}\relax
\EndOfBibitem
\bibitem[Ai \latin{et~al.}(2019)Ai, Kou, Hu, Wang, Krasheninnikov, Sun, and
  Shen]{Ai2019}
Ai,~W.; Kou,~L.; Hu,~X.; Wang,~Y.; Krasheninnikov,~A.~V.; Sun,~L.; Shen,~X.
  Enhanced Sensitivity of {MoSe}$_2$ Monolayer for Gas Adsorption Induced by
  Electric Field. \emph{J. Phys-Condens. Mat.} \textbf{2019}, \emph{31},
  445301\relax
\mciteBstWouldAddEndPuncttrue
\mciteSetBstMidEndSepPunct{\mcitedefaultmidpunct}
{\mcitedefaultendpunct}{\mcitedefaultseppunct}\relax
\EndOfBibitem
\bibitem[{dos Santos} \latin{et~al.}(2021){dos Santos}, {da Cunha}, Giozza, {de
  Sousa Júnior}, Roncaratti, and {Ribeiro Júnior}]{Santos2020}
{dos Santos},~R.~M.; {da Cunha},~W.~F.; Giozza,~W.~F.; {de Sousa
  Júnior},~R.~T.; Roncaratti,~L.~F.; {Ribeiro Júnior},~L.~A. Electronic and
  Structural Properties of Janus {MoSSe/MoX$_2$} ({X=S,Se}) in-plane
  Heterojunctions: A {DFT} Study. \emph{Chem. Phys. Lett} \textbf{2021},
  \emph{771}, 138495\relax
\mciteBstWouldAddEndPuncttrue
\mciteSetBstMidEndSepPunct{\mcitedefaultmidpunct}
{\mcitedefaultendpunct}{\mcitedefaultseppunct}\relax
\EndOfBibitem
\bibitem[Liang \latin{et~al.}(2021)Liang, Zhang, Zhao, Liu, and Wee]{Liang2021}
Liang,~Q.; Zhang,~Q.; Zhao,~X.; Liu,~M.; Wee,~A. T.~S. Defect Engineering of
  Two-Dimensional Transition-Metal Dichalcogenides: Applications, Challenges,
  and Opportunities. \emph{ACS Nano} \textbf{2021}, \emph{15}, 2165--2181,
  PMID: 33449623\relax
\mciteBstWouldAddEndPuncttrue
\mciteSetBstMidEndSepPunct{\mcitedefaultmidpunct}
{\mcitedefaultendpunct}{\mcitedefaultseppunct}\relax
\EndOfBibitem
\bibitem[Koós \latin{et~al.}(2019)Koós, Vancsó, Szendrő, Dobrik,
  Antognini~Silva, Popov, Sorokin, Henrard, Hwang, Biró, and
  Tapasztó]{Koos2019}
Koós,~A.~A.; Vancsó,~P.; Szendrő,~M.; Dobrik,~G.; Antognini~Silva,~D.;
  Popov,~Z.~I.; Sorokin,~P.~B.; Henrard,~L.; Hwang,~C.; Biró,~L.~P.
  \latin{et~al.}  Influence of Native Defects on the Electronic and Magnetic
  Properties of {CVD} Grown {MoSe$_2$} Single Layers. \emph{J. Phys. Chem. C}
  \textbf{2019}, \emph{123}, 24855--24864\relax
\mciteBstWouldAddEndPuncttrue
\mciteSetBstMidEndSepPunct{\mcitedefaultmidpunct}
{\mcitedefaultendpunct}{\mcitedefaultseppunct}\relax
\EndOfBibitem
\bibitem[Zhou \latin{et~al.}(2021)Zhou, Dong, Tan, and Tang]{Zhou2021HER}
Zhou,~W.; Dong,~L.; Tan,~L.; Tang,~Q. Understanding the Air Stability of
  Defective {MoS$_2$} and the Oxidation Effect on the Surface {HER} Activity.
  \emph{J. Phys-Condens Mat} \textbf{2021}, \emph{33}, 395002\relax
\mciteBstWouldAddEndPuncttrue
\mciteSetBstMidEndSepPunct{\mcitedefaultmidpunct}
{\mcitedefaultendpunct}{\mcitedefaultseppunct}\relax
\EndOfBibitem
\bibitem[Lunardon \latin{et~al.}(2023)Lunardon, Kosmala, Ghorbani-Asl,
  Krasheninnikov, Kolekar, Durante, Batzill, Agnoli, and
  Granozzi]{Lunardon2023}
Lunardon,~M.; Kosmala,~T.; Ghorbani-Asl,~M.; Krasheninnikov,~A.~V.;
  Kolekar,~S.; Durante,~C.; Batzill,~M.; Agnoli,~S.; Granozzi,~G. Catalytic
  Activity of Defect-Engineered Transition Me tal Dichalcogenides Mapped with
  Atomic-Scale Precision by Electrochemical Scanning Tunneling Microscopy.
  \emph{ACS Energy Lett.} \textbf{2023}, \emph{8}, 972--980\relax
\mciteBstWouldAddEndPuncttrue
\mciteSetBstMidEndSepPunct{\mcitedefaultmidpunct}
{\mcitedefaultendpunct}{\mcitedefaultseppunct}\relax
\EndOfBibitem
\bibitem[Feng \latin{et~al.}(2016)Feng, Chen, Qian, Xu, Feng, Yu, Zhang, Chen,
  Li, Li, Sun, Zhang, Liu, Pang, and Zhang]{Feng2016}
Feng,~Z.; Chen,~B.; Qian,~S.; Xu,~L.; Feng,~L.; Yu,~Y.; Zhang,~R.; Chen,~J.;
  Li,~Q.; Li,~Q. \latin{et~al.}  Chemical sensing by band modulation of a black
  phosphorus/molybdenum diselenide van der Waals hetero-structure. \emph{2D
  Mater.} \textbf{2016}, \emph{3}, 035021\relax
\mciteBstWouldAddEndPuncttrue
\mciteSetBstMidEndSepPunct{\mcitedefaultmidpunct}
{\mcitedefaultendpunct}{\mcitedefaultseppunct}\relax
\EndOfBibitem
\bibitem[Yang \latin{et~al.}(2020)Yang, Zhang, and Wang]{YANG2020127369}
Yang,~Z.; Zhang,~D.; Wang,~D. Carbon monoxide gas sensing properties of
  metal-organic frameworks-derived tin dioxide nanoparticles/molybdenum
  diselenide nanoflowers. \emph{Sensor Actuat B-Chem} \textbf{2020},
  \emph{304}, 127369\relax
\mciteBstWouldAddEndPuncttrue
\mciteSetBstMidEndSepPunct{\mcitedefaultmidpunct}
{\mcitedefaultendpunct}{\mcitedefaultseppunct}\relax
\EndOfBibitem
\bibitem[Zhou \latin{et~al.}(2021)Zhou, Mi, Jin, Li, and Zhang]{Zhou2021flower}
Zhou,~L.; Mi,~Q.; Jin,~Y.; Li,~T.; Zhang,~D. Construction of MoO3/MoSe$_2$
  nanocomposite-based gas sensor for low detection limit trimethylamine sensing
  at room temperature. \emph{J. Mater. Sci.: Mater. Electron.} \textbf{2021},
  \emph{32}, 17301--17310\relax
\mciteBstWouldAddEndPuncttrue
\mciteSetBstMidEndSepPunct{\mcitedefaultmidpunct}
{\mcitedefaultendpunct}{\mcitedefaultseppunct}\relax
\EndOfBibitem
\bibitem[Jaiswal \latin{et~al.}(2022)Jaiswal, Das, Chetry, Kumar, and
  Chandra]{Jaiswal2022}
Jaiswal,~J.; Das,~A.; Chetry,~V.; Kumar,~S.; Chandra,~R. NO2 sensors based on
  crystalline MoSe$_2$ porous nanowall thin films with vertically aligned
  molecular layers prepared by sputtering. \emph{Sens. Actuator B-Chem.}
  \textbf{2022}, \emph{359}, 131552\relax
\mciteBstWouldAddEndPuncttrue
\mciteSetBstMidEndSepPunct{\mcitedefaultmidpunct}
{\mcitedefaultendpunct}{\mcitedefaultseppunct}\relax
\EndOfBibitem
\bibitem[Liu \latin{et~al.}(2013)Liu, Ahsan, Khitun, Lake, and
  Balandin]{Guanxiong2013}
Liu,~G.; Ahsan,~S.; Khitun,~A.~G.; Lake,~R.~K.; Balandin,~A.~A. Graphene-based
  non-Boolean logic circuits. \emph{J. Appl. Phys.} \textbf{2013}, \emph{114},
  154310\relax
\mciteBstWouldAddEndPuncttrue
\mciteSetBstMidEndSepPunct{\mcitedefaultmidpunct}
{\mcitedefaultendpunct}{\mcitedefaultseppunct}\relax
\EndOfBibitem
\bibitem[Shu \latin{et~al.}(2017)Shu, Zhou, Li, Cao, and Chen]{Shu2017}
Shu,~H.; Zhou,~D.; Li,~F.; Cao,~D.; Chen,~X. Defect Engineering in {MoSe$_2$}
  for the Hydrogen Evolution Reaction: From Point Defects to Edges. \emph{ACS
  Appl. Mater. Inter.} \textbf{2017}, \emph{9}, 42688--42698, PMID:
  29152972\relax
\mciteBstWouldAddEndPuncttrue
\mciteSetBstMidEndSepPunct{\mcitedefaultmidpunct}
{\mcitedefaultendpunct}{\mcitedefaultseppunct}\relax
\EndOfBibitem
\bibitem[Lee \latin{et~al.}(2018)Lee, Kang, Yim, Kim, Jang, Kang, and
  Han]{Lee2018}
Lee,~J.; Kang,~S.; Yim,~K.; Kim,~K.~Y.; Jang,~H.~W.; Kang,~Y.; Han,~S. Hydrogen
  Evolution Reaction at Anion Vacancy of Two-Dimensional Transition-Metal
  Dichalcogenides: Ab Initio Computational Screening. \emph{J. Phys. Chem.
  Lett.} \textbf{2018}, \emph{9}, 2049--2055, PMID: 29621882\relax
\mciteBstWouldAddEndPuncttrue
\mciteSetBstMidEndSepPunct{\mcitedefaultmidpunct}
{\mcitedefaultendpunct}{\mcitedefaultseppunct}\relax
\EndOfBibitem
\bibitem[Jain \latin{et~al.}(2020)Jain, Sadan, and Ramasubramaniam]{Jain2020}
Jain,~A.; Sadan,~M.~B.; Ramasubramaniam,~A. Promoting Active Sites for Hydrogen
  Evolution in {MoSe}$_2$ via Transition-Metal Doping. \emph{J. Phys. Chem. C}
  \textbf{2020}, \emph{124}, 12324--12336\relax
\mciteBstWouldAddEndPuncttrue
\mciteSetBstMidEndSepPunct{\mcitedefaultmidpunct}
{\mcitedefaultendpunct}{\mcitedefaultseppunct}\relax
\EndOfBibitem
\bibitem[Yi \latin{et~al.}(2019)Yi, Li, Gong, She, Song, Xu, Deng, Yuan, Xu,
  and Li]{Yi2019}
Yi,~J.; Li,~H.; Gong,~Y.; She,~X.; Song,~Y.; Xu,~Y.; Deng,~J.; Yuan,~S.;
  Xu,~H.; Li,~H. Phase and interlayer effect of transition metal dichalcogenide
  cocatalyst toward photocatalytic hydrogen evolution: The case of MoSe$_2$.
  \emph{Appl. Catal. B: Environ} \textbf{2019}, \emph{243}, 330--336\relax
\mciteBstWouldAddEndPuncttrue
\mciteSetBstMidEndSepPunct{\mcitedefaultmidpunct}
{\mcitedefaultendpunct}{\mcitedefaultseppunct}\relax
\EndOfBibitem
\bibitem[German and Gebauer(2020)German, and Gebauer]{German2020}
German,~E.; Gebauer,~R. Why are {MoS$_2$} monolayers not a good catalyst for
  the oxygen evolution reaction? \emph{Appl. Surf. Sci.} \textbf{2020},
  \emph{528}, 146591\relax
\mciteBstWouldAddEndPuncttrue
\mciteSetBstMidEndSepPunct{\mcitedefaultmidpunct}
{\mcitedefaultendpunct}{\mcitedefaultseppunct}\relax
\EndOfBibitem
\bibitem[Karmodak and Andreussi(2021)Karmodak, and Andreussi]{Karmodak2021}
Karmodak,~N.; Andreussi,~O. Oxygen Evolution on {MoS$_2$} Edges: Activation
  through Surface Oxidation. \emph{J. Phys. Chem. C} \textbf{2021}, \emph{125},
  10397--10405\relax
\mciteBstWouldAddEndPuncttrue
\mciteSetBstMidEndSepPunct{\mcitedefaultmidpunct}
{\mcitedefaultendpunct}{\mcitedefaultseppunct}\relax
\EndOfBibitem
\bibitem[Mohanty \latin{et~al.}(2018)Mohanty, Ghorbani-Asl, Kretschmer, Ghosh,
  Guha, Panda, Jena, Krasheninnikov, and Jena]{Mohanty2018}
Mohanty,~B.; Ghorbani-Asl,~M.; Kretschmer,~S.; Ghosh,~A.; Guha,~P.;
  Panda,~S.~K.; Jena,~B.; Krasheninnikov,~A.~V.; Jena,~B.~K. {MoS$_2$} Quantum
  Dots as Efficient Catalyst Materials for the Oxygen Evolution Reaction.
  \emph{ACS Catal.} \textbf{2018}, \emph{8}, 1683--1689\relax
\mciteBstWouldAddEndPuncttrue
\mciteSetBstMidEndSepPunct{\mcitedefaultmidpunct}
{\mcitedefaultendpunct}{\mcitedefaultseppunct}\relax
\EndOfBibitem
\bibitem[Chan \latin{et~al.}(2014)Chan, Tsai, Hansen, and Nørskov]{chan2014}
Chan,~K.; Tsai,~C.; Hansen,~H.~A.; Nørskov,~J.~K. Molybdenum Sulfides and
  Selenides as Possible Electrocatalysts for {CO$_2$} Reduction.
  \emph{ChemCatChem} \textbf{2014}, \emph{6}, 1899--1905\relax
\mciteBstWouldAddEndPuncttrue
\mciteSetBstMidEndSepPunct{\mcitedefaultmidpunct}
{\mcitedefaultendpunct}{\mcitedefaultseppunct}\relax
\EndOfBibitem
\bibitem[Aguilar \latin{et~al.}(2020)Aguilar, Atilhan, and
  Aparicio]{AGUILAR2020147611}
Aguilar,~N.; Atilhan,~M.; Aparicio,~S. Single atom transition metals on
  {MoS}$_2$ monolayer and their use as catalysts for {CO$_2$} activation.
  \emph{Appl. Surf. Sci.} \textbf{2020}, \emph{534}, 147611\relax
\mciteBstWouldAddEndPuncttrue
\mciteSetBstMidEndSepPunct{\mcitedefaultmidpunct}
{\mcitedefaultendpunct}{\mcitedefaultseppunct}\relax
\EndOfBibitem
\bibitem[Kang \latin{et~al.}(2019)Kang, Han, and Kang]{Kang2019COr}
Kang,~S.; Han,~S.; Kang,~Y. Unveiling Electrochemical Reaction Pathways of
  {CO$_2$} Reduction to {CN} Species at {S-Vacancies} of {MoS$_2$}.
  \emph{ChemSusChem} \textbf{2019}, \emph{12}, 2671--2678\relax
\mciteBstWouldAddEndPuncttrue
\mciteSetBstMidEndSepPunct{\mcitedefaultmidpunct}
{\mcitedefaultendpunct}{\mcitedefaultseppunct}\relax
\EndOfBibitem
\bibitem[Ye \latin{et~al.}(2021)Ye, Rao, and Yan]{Ye2021}
Ye,~J.; Rao,~D.; Yan,~X. Regulating the electronic properties of {MoSe$_2$} to
  improve its {CO$_2$} electrocatalytic reduction performance via atomic
  doping. \emph{New J. Chem.} \textbf{2021}, \emph{45}, 5350--5356\relax
\mciteBstWouldAddEndPuncttrue
\mciteSetBstMidEndSepPunct{\mcitedefaultmidpunct}
{\mcitedefaultendpunct}{\mcitedefaultseppunct}\relax
\EndOfBibitem
\bibitem[Ren \latin{et~al.}(2022)Ren, Sun, Qi, and Zhao]{Ren2021}
Ren,~Y.; Sun,~X.; Qi,~K.; Zhao,~Z. Single atom supported on MoS$_2$ as
  efficient electrocatalysts for the {CO$_2$} reduction reaction: A DFT study.
  \emph{Appl. Surf. Sci.} \textbf{2022}, \emph{602}, 154211\relax
\mciteBstWouldAddEndPuncttrue
\mciteSetBstMidEndSepPunct{\mcitedefaultmidpunct}
{\mcitedefaultendpunct}{\mcitedefaultseppunct}\relax
\EndOfBibitem
\bibitem[Ramzan \latin{et~al.}(2024)Ramzan, Favre, Steinmann, Le~Bahers, and
  Raybaud]{Ramzan2024}
Ramzan,~M.~A.; Favre,~R.; Steinmann,~S.~N.; Le~Bahers,~T.; Raybaud,~P.
  Electrocatalytic Reduction Mechanisms of {CO$_2$} on MoS$_2$ Edges Using
  Grand-Canonical DFT: From {CO$_2$} Adsorption to HCOOH or CO. \emph{J. Phys.
  Chem. C} \textbf{2024}, \emph{128}, 10025--10034\relax
\mciteBstWouldAddEndPuncttrue
\mciteSetBstMidEndSepPunct{\mcitedefaultmidpunct}
{\mcitedefaultendpunct}{\mcitedefaultseppunct}\relax
\EndOfBibitem
\bibitem[Zhang \latin{et~al.}(2020)Zhang, Ling, Zang, Li, Huang, Li, Yan, Kou,
  Liu, Wang, and Yang]{Zhang2020}
Zhang,~J.; Ling,~C.; Zang,~W.; Li,~X.; Huang,~S.; Li,~X.~L.; Yan,~D.; Kou,~Z.;
  Liu,~L.; Wang,~J. \latin{et~al.}  Boosted electrochemical ammonia synthesis
  by high-percentage metallic transition metal dichalcogenide quantum dots.
  \emph{Nanoscale} \textbf{2020}, \emph{12}, 10964--10971\relax
\mciteBstWouldAddEndPuncttrue
\mciteSetBstMidEndSepPunct{\mcitedefaultmidpunct}
{\mcitedefaultendpunct}{\mcitedefaultseppunct}\relax
\EndOfBibitem
\bibitem[Qu \latin{et~al.}(2020)Qu, He, Su, Zhang, and Su]{Qu2020}
Qu,~H.; He,~S.; Su,~Y.; Zhang,~Y.; Su,~H. MoSe$_2$: a promising non-noble metal
  catalyst for direct ethanol synthesis from syngas. \emph{Fuel} \textbf{2020},
  \emph{281}, 118760\relax
\mciteBstWouldAddEndPuncttrue
\mciteSetBstMidEndSepPunct{\mcitedefaultmidpunct}
{\mcitedefaultendpunct}{\mcitedefaultseppunct}\relax
\EndOfBibitem
\bibitem[Ma \latin{et~al.}(2015)Ma, Tang, Yang, Zeng, He, and Lu]{MA2015}
Ma,~D.; Tang,~Y.; Yang,~G.; Zeng,~J.; He,~C.; Lu,~Z. CO catalytic oxidation on
  iron-embedded monolayer MoS$_2$. \emph{Appl. Surf. Sci.} \textbf{2015},
  \emph{328}, 71--77\relax
\mciteBstWouldAddEndPuncttrue
\mciteSetBstMidEndSepPunct{\mcitedefaultmidpunct}
{\mcitedefaultendpunct}{\mcitedefaultseppunct}\relax
\EndOfBibitem
\bibitem[Zhu \latin{et~al.}(2019)Zhu, Zhao, Shi, Ren, Zhao, Shang, Xue, Guo,
  Duan, He, Guo, and Li]{Zhu2019}
Zhu,~Y.; Zhao,~K.; Shi,~J.; Ren,~X.; Zhao,~X.; Shang,~Y.; Xue,~X.; Guo,~H.;
  Duan,~X.; He,~H. \latin{et~al.}  Strain Engineering of a Defect-Free,
  Single-Layer MoS$_2$ Substrate for Highly Efficient Single-Atom Catalysis of
  CO Oxidation. \emph{ACS Appl. Mater. Interfaces} \textbf{2019}, \emph{11},
  32887--32894, PMID: 31429270\relax
\mciteBstWouldAddEndPuncttrue
\mciteSetBstMidEndSepPunct{\mcitedefaultmidpunct}
{\mcitedefaultendpunct}{\mcitedefaultseppunct}\relax
\EndOfBibitem
\bibitem[Zhang \latin{et~al.}(2019)Zhang, Chang, Yang, and Wang]{Zhang2020COox}
Zhang,~X.; Chang,~Q.; Yang,~Z.; Wang,~W. Surface vacancy on PtTe2 for promoting
  CO oxidation through efficiently activating O2. \emph{J. Phys-Condens. Mat.}
  \textbf{2019}, \emph{32}, 025201\relax
\mciteBstWouldAddEndPuncttrue
\mciteSetBstMidEndSepPunct{\mcitedefaultmidpunct}
{\mcitedefaultendpunct}{\mcitedefaultseppunct}\relax
\EndOfBibitem
\bibitem[Guo \latin{et~al.}(2021)Guo, Wu, Zhong, Zhang, Shen, and Yu]{Guo2021}
Guo,~J.-X.; Wu,~S.-Y.; Zhong,~S.-Y.; Zhang,~G.-J.; Shen,~G.-Q.; Yu,~X.-Y. Janus
  {WSSe} monolayer adsorbed with transition-metal atoms ({Fe}, {Co} and {Ni}):
  excellent performance for gas sensing and {CO} catalytic oxidation.
  \emph{Appl. Surf. Sci.} \textbf{2021}, \emph{565}, 150558\relax
\mciteBstWouldAddEndPuncttrue
\mciteSetBstMidEndSepPunct{\mcitedefaultmidpunct}
{\mcitedefaultendpunct}{\mcitedefaultseppunct}\relax
\EndOfBibitem
\bibitem[Bomb\'{\i}n \latin{et~al.}(2022)Bomb\'{\i}n, Alducin, and
  Juaristi]{bombin2022}
Bomb\'{\i}n,~R.; Alducin,~M.; Juaristi,~J. I.~n. Adsorption and dissociation of
  diatomic molecules on monolayer $1H\text{\ensuremath{-}}{\mathrm{MoSe}}_{2}$.
  \emph{Phys. Rev. B} \textbf{2022}, \emph{105}, 035404\relax
\mciteBstWouldAddEndPuncttrue
\mciteSetBstMidEndSepPunct{\mcitedefaultmidpunct}
{\mcitedefaultendpunct}{\mcitedefaultseppunct}\relax
\EndOfBibitem
\bibitem[Kresse and Furthm\"uller(1996)Kresse, and Furthm\"uller]{Kresse1996}
Kresse,~G.; Furthm\"uller,~J. Efficient iterative schemes for ab initio
  total-energy calculations using a plane-wave basis set. \emph{Phys. Rev. B}
  \textbf{1996}, \emph{54}, 11169--11186\relax
\mciteBstWouldAddEndPuncttrue
\mciteSetBstMidEndSepPunct{\mcitedefaultmidpunct}
{\mcitedefaultendpunct}{\mcitedefaultseppunct}\relax
\EndOfBibitem
\bibitem[Dion \latin{et~al.}(2004)Dion, Rydberg, Schr\"oder, Langreth, and
  Lundqvist]{Dion2004}
Dion,~M.; Rydberg,~H.; Schr\"oder,~E.; Langreth,~D.~C.; Lundqvist,~B.~I. Van
  der {Waals} Density Functional for General Geometries. \emph{Phys. Rev.
  Lett.} \textbf{2004}, \emph{92}, 246401\relax
\mciteBstWouldAddEndPuncttrue
\mciteSetBstMidEndSepPunct{\mcitedefaultmidpunct}
{\mcitedefaultendpunct}{\mcitedefaultseppunct}\relax
\EndOfBibitem
\bibitem[Bl\"ochl(1994)]{blockl1994}
Bl\"ochl,~P.~E. Projector augmented-wave method. \emph{Phys. Rev. B}
  \textbf{1994}, \emph{50}, 17953--17979\relax
\mciteBstWouldAddEndPuncttrue
\mciteSetBstMidEndSepPunct{\mcitedefaultmidpunct}
{\mcitedefaultendpunct}{\mcitedefaultseppunct}\relax
\EndOfBibitem
\bibitem[Kresse and Joubert(1999)Kresse, and Joubert]{Kresse1999}
Kresse,~G.; Joubert,~D. From ultrasoft pseudopotentials to the projector
  augmented-wave method. \emph{Phys. Rev. B} \textbf{1999}, \emph{59},
  1758--1775\relax
\mciteBstWouldAddEndPuncttrue
\mciteSetBstMidEndSepPunct{\mcitedefaultmidpunct}
{\mcitedefaultendpunct}{\mcitedefaultseppunct}\relax
\EndOfBibitem
\bibitem[Monkhorst and Pack(1976)Monkhorst, and Pack]{Monkhorst1976}
Monkhorst,~H.~J.; Pack,~J.~D. Special points for Brillouin-zone integrations.
  \emph{Phys. Rev. B} \textbf{1976}, \emph{13}, 5188--5192\relax
\mciteBstWouldAddEndPuncttrue
\mciteSetBstMidEndSepPunct{\mcitedefaultmidpunct}
{\mcitedefaultendpunct}{\mcitedefaultseppunct}\relax
\EndOfBibitem
\bibitem[Grimme \latin{et~al.}(2010)Grimme, Antony, Ehrlich, and
  Krieg]{Grime2010}
Grimme,~S.; Antony,~J.; Ehrlich,~S.; Krieg,~H. A consistent and accurate ab
  initio parametrization of density functional dispersion correction ({DFT-D})
  for the 94 elements {H-Pu}. \emph{J. Chem. Phys.} \textbf{2010}, \emph{132},
  154104\relax
\mciteBstWouldAddEndPuncttrue
\mciteSetBstMidEndSepPunct{\mcitedefaultmidpunct}
{\mcitedefaultendpunct}{\mcitedefaultseppunct}\relax
\EndOfBibitem
\bibitem[Nos\'e(1984)]{Nose84}
Nos\'e,~S. A Unified Formulation of the Constant Temperature Molecular Dynamics
  Methods. \emph{J. Chem. Phys.} \textbf{1984}, \emph{81}, 511--519\relax
\mciteBstWouldAddEndPuncttrue
\mciteSetBstMidEndSepPunct{\mcitedefaultmidpunct}
{\mcitedefaultendpunct}{\mcitedefaultseppunct}\relax
\EndOfBibitem
\bibitem[Novko \latin{et~al.}(2017)Novko, Lon\ifmmode \check{c}\else
  \v{c}\fi{}ari\ifmmode~\acute{c}\else \'{c}\fi{}, Blanco-Rey, Juaristi, and
  Alducin]{Novko2017}
Novko,~D.; Lon\ifmmode \check{c}\else \v{c}\fi{}ari\ifmmode~\acute{c}\else
  \'{c}\fi{},~I.; Blanco-Rey,~M.; Juaristi,~J.~I.; Alducin,~M. Energy loss and
  surface temperature effects in ab initio molecular dynamics simulations: N
  adsorption on Ag(111) as a case study. \emph{Phys. Rev. B} \textbf{2017},
  \emph{96}, 085437\relax
\mciteBstWouldAddEndPuncttrue
\mciteSetBstMidEndSepPunct{\mcitedefaultmidpunct}
{\mcitedefaultendpunct}{\mcitedefaultseppunct}\relax
\EndOfBibitem
\bibitem[Gro\ss{} and Dianat(2007)Gro\ss{}, and Dianat]{Gross07}
Gro\ss{},~A.; Dianat,~A. Hydrogen Dissociation Dynamics on Precovered Pd
  Surfaces: Langmuir is Still Right. \emph{Phys. Rev. Lett.} \textbf{2007},
  \emph{98}, 206107\relax
\mciteBstWouldAddEndPuncttrue
\mciteSetBstMidEndSepPunct{\mcitedefaultmidpunct}
{\mcitedefaultendpunct}{\mcitedefaultseppunct}\relax
\EndOfBibitem
\bibitem[Nattino \latin{et~al.}(2012)Nattino, D\'{\i}az, Jackson, and
  Kroes]{Nattino12}
Nattino,~F.; D\'{\i}az,~C.; Jackson,~B.; Kroes,~G.-J. Effect of Surface Motion
  on the Rotational Quadrupole Alignment Parameter of ${\mathbf{D}}_{2}$
  Reacting on Cu(111). \emph{Phys. Rev. Lett.} \textbf{2012}, \emph{108},
  236104\relax
\mciteBstWouldAddEndPuncttrue
\mciteSetBstMidEndSepPunct{\mcitedefaultmidpunct}
{\mcitedefaultendpunct}{\mcitedefaultseppunct}\relax
\EndOfBibitem
\bibitem[Nattino \latin{et~al.}(2014)Nattino, Ueta, Chadwick, van Reijzen,
  Beck, Jackson, van Hemert, and Kroes]{Nattino14}
Nattino,~F.; Ueta,~H.; Chadwick,~H.; van Reijzen,~M.~E.; Beck,~R.~D.;
  Jackson,~B.; van Hemert,~M.~C.; Kroes,~G.-J. Ab Initio Molecular Dynamics
  Calculations versus Quantum-State-Resolved Experiments on CHD$_3$ + Pt(111):
  New Insights into a Prototypical Gas Surface Reaction. \emph{J. Phys. Chem.
  Lett.} \textbf{2014}, \emph{5}, 1294--1299, PMID: 26269970\relax
\mciteBstWouldAddEndPuncttrue
\mciteSetBstMidEndSepPunct{\mcitedefaultmidpunct}
{\mcitedefaultendpunct}{\mcitedefaultseppunct}\relax
\EndOfBibitem
\bibitem[Kolb and Guo(2016)Kolb, and Guo]{Kolb16}
Kolb,~B.; Guo,~H. Communication: Energy transfer and reaction dynamics for DCl
  scattering on Au(111): An ab initio molecular dynamics study. \emph{J. Chem.
  Phys.} \textbf{2016}, \emph{145}, 011102\relax
\mciteBstWouldAddEndPuncttrue
\mciteSetBstMidEndSepPunct{\mcitedefaultmidpunct}
{\mcitedefaultendpunct}{\mcitedefaultseppunct}\relax
\EndOfBibitem
\bibitem[Zhou \latin{et~al.}(2017)Zhou, Kolb, Luo, Guo, and Jiang]{Zhou17}
Zhou,~X.; Kolb,~B.; Luo,~X.; Guo,~H.; Jiang,~B. Ab Initio Molecular Dynamics
  Study of Dissociative Chemisorption and Scattering of {CO$_2$} on Ni(100):
  Reactivity, Energy Transfer, Steering Dynamics, and Lattice Effects. \emph{J.
  Phys. Chem. C} \textbf{2017}, \emph{121}, 5594--5602\relax
\mciteBstWouldAddEndPuncttrue
\mciteSetBstMidEndSepPunct{\mcitedefaultmidpunct}
{\mcitedefaultendpunct}{\mcitedefaultseppunct}\relax
\EndOfBibitem
\bibitem[Zhou \latin{et~al.}(2018)Zhou, Jiang, Alducin, and Guo]{Zhoujcp18b}
Zhou,~L.; Jiang,~B.; Alducin,~M.; Guo,~H. Communication: Fingerprints of
  reaction mechanisms in product distributions: Eley-Rideal-type reactions
  between D and CD3/Cu(111). \emph{J. Chem. Phys.} \textbf{2018}, \emph{149},
  031101\relax
\mciteBstWouldAddEndPuncttrue
\mciteSetBstMidEndSepPunct{\mcitedefaultmidpunct}
{\mcitedefaultendpunct}{\mcitedefaultseppunct}\relax
\EndOfBibitem
\bibitem[F\"uchsel \latin{et~al.}(2019)F\"uchsel, Zhou, Jiang, Juaristi,
  Alducin, Guo, and Kroes]{Fuchseljpcc19}
F\"uchsel,~G.; Zhou,~X.; Jiang,~B.; Juaristi,~J.~I.; Alducin,~M.; Guo,~H.;
  Kroes,~G.-J. Reactive and Nonreactive Scattering of HCl from Au(111): An Ab
  Initio Molecular Dynamics Study. \emph{J. Phys. Chem. C} \textbf{2019},
  \emph{123}, 2287--2299\relax
\mciteBstWouldAddEndPuncttrue
\mciteSetBstMidEndSepPunct{\mcitedefaultmidpunct}
{\mcitedefaultendpunct}{\mcitedefaultseppunct}\relax
\EndOfBibitem
\bibitem[Rivero~Santamar\'{\i}a \latin{et~al.}(2019)Rivero~Santamar\'{\i}a,
  Alducin, D\'{\i}ez Mui\~no, and Juaristi]{santamaria2019}
Rivero~Santamar\'{\i}a,~A.; Alducin,~M.; D\'{\i}ez Mui\~no,~R.; Juaristi,~J.
  I.~n. Ab Initio Molecular Dynamics Study of Alignment-Resolved O2 Scattering
  from Highly Oriented Pyrolytic Graphite. \emph{J. Phys. Chem. C}
  \textbf{2019}, \emph{123}, 31094--31102\relax
\mciteBstWouldAddEndPuncttrue
\mciteSetBstMidEndSepPunct{\mcitedefaultmidpunct}
{\mcitedefaultendpunct}{\mcitedefaultseppunct}\relax
\EndOfBibitem
\bibitem[Rivero~Santamar\'{\i}a \latin{et~al.}(2021)Rivero~Santamar\'{\i}a,
  Ramos, Alducin, Busnengo, D\'{\i}ez Mui\~no, and Juaristi]{santamaria2021}
Rivero~Santamar\'{\i}a,~A.; Ramos,~M.; Alducin,~M.; Busnengo,~H.~F.; D\'{\i}ez
  Mui\~no,~R.; Juaristi,~J. I.~n. High-Dimensional Atomistic Neural Network
  Potential to Study the Alignment-Resolved O2 Scattering from Highly Oriented
  Pyrolytic Graphite. \emph{J. Phys. Chem. A} \textbf{2021}, \emph{125},
  2588--2600, PMID: 33734696\relax
\mciteBstWouldAddEndPuncttrue
\mciteSetBstMidEndSepPunct{\mcitedefaultmidpunct}
{\mcitedefaultendpunct}{\mcitedefaultseppunct}\relax
\EndOfBibitem
\bibitem[Granja-DelR\'{\i}o \latin{et~al.}(2021)Granja-DelR\'{\i}o, Alducin,
  naki Juaristi, L\'{o}pez, and Alonso]{GRANJADELRIO2021}
Granja-DelR\'{\i}o,~A.; Alducin,~M.; naki Juaristi,~J.~I.; L\'{o}pez,~M.~J.;
  Alonso,~J.~A. Absence of spillover of hydrogen adsorbed on small palladium
  clusters anchored to graphene vacancies. \emph{Appl. Surf. Sci.}
  \textbf{2021}, \emph{559}, 149835\relax
\mciteBstWouldAddEndPuncttrue
\mciteSetBstMidEndSepPunct{\mcitedefaultmidpunct}
{\mcitedefaultendpunct}{\mcitedefaultseppunct}\relax
\EndOfBibitem
\bibitem[Quattrucci and Jackson(2005)Quattrucci, and Jackson]{Quattrucci2005}
Quattrucci,~J.~G.; Jackson,~B. {Quasiclassical study of Eley–Rideal and hot
  atom reactions of H atoms with Cl adsorbed on a Au(111) surface}. \emph{J.
  Chem. Phys.} \textbf{2005}, \emph{122}, 074705\relax
\mciteBstWouldAddEndPuncttrue
\mciteSetBstMidEndSepPunct{\mcitedefaultmidpunct}
{\mcitedefaultendpunct}{\mcitedefaultseppunct}\relax
\EndOfBibitem
\bibitem[Zhou \latin{et~al.}(2018)Zhou, Zhou, Alducin, Zhang, Jiang, and
  Guo]{Zhou2018}
Zhou,~L.; Zhou,~X.; Alducin,~M.; Zhang,~L.; Jiang,~B.; Guo,~H. {Ab initio
  molecular dynamics study of the Eley-Rideal reaction of H + Cl–Au(111) →
  HCl + Au(111): Impact of energy dissipation to surface phonons and
  electron-hole pairs}. \emph{J. Chem. Phys.} \textbf{2018}, \emph{148},
  014702\relax
\mciteBstWouldAddEndPuncttrue
\mciteSetBstMidEndSepPunct{\mcitedefaultmidpunct}
{\mcitedefaultendpunct}{\mcitedefaultseppunct}\relax
\EndOfBibitem
\bibitem[Chen \latin{et~al.}(2019)Chen, Zhou, and Jiang]{Chen2019}
Chen,~J.; Zhou,~X.; Jiang,~B. {Eley Rideal recombination of hydrogen atoms on
  Cu(111): Quantitative role of electronic excitation in cross sections and
  product distributions}. \emph{J. Chem. Phys.} \textbf{2019}, \emph{150},
  061101\relax
\mciteBstWouldAddEndPuncttrue
\mciteSetBstMidEndSepPunct{\mcitedefaultmidpunct}
{\mcitedefaultendpunct}{\mcitedefaultseppunct}\relax
\EndOfBibitem
\bibitem[Zhu \latin{et~al.}(2023)Zhu, Hu, Chen, and Jiang]{Zhu2023}
Zhu,~L.; Hu,~C.; Chen,~J.; Jiang,~B. Investigating the Eley–Rideal
  recombination of hydrogen atoms on Cu (111) via a high-dimensional neural
  network potential energy surface. \emph{Phys. Chem. Chem. Phys.}
  \textbf{2023}, \emph{25}, 5479--5488\relax
\mciteBstWouldAddEndPuncttrue
\mciteSetBstMidEndSepPunct{\mcitedefaultmidpunct}
{\mcitedefaultendpunct}{\mcitedefaultseppunct}\relax
\EndOfBibitem
\bibitem[Blanco-Rey \latin{et~al.}(2013)Blanco-Rey, Díaz, Bocan, Díez~Muiño,
  Alducin, and Juaristi]{Blanco-Rey2013}
Blanco-Rey,~M.; Díaz,~E.; Bocan,~G.~A.; Díez~Muiño,~R.; Alducin,~M.;
  Juaristi,~J.~I. Efficient N2 Formation on Ag(111) by Eley–Rideal
  Recombination of Hyperthermal Atoms. \emph{J. Phys. Chem. Lett.}
  \textbf{2013}, \emph{4}, 3704--3709\relax
\mciteBstWouldAddEndPuncttrue
\mciteSetBstMidEndSepPunct{\mcitedefaultmidpunct}
{\mcitedefaultendpunct}{\mcitedefaultseppunct}\relax
\EndOfBibitem
\bibitem[Juaristi \latin{et~al.}(2015)Juaristi, Díaz, Bocan, {Díez Muiño},
  Alducin, and Blanco-Rey]{Juaristi2015}
Juaristi,~J.~I.; Díaz,~E.; Bocan,~G.; {Díez Muiño},~R.; Alducin,~M.;
  Blanco-Rey,~M. Angular distributions and rovibrational excitation of N2
  molecules recombined on N-covered Ag(111) by the Eley–Rideal mechanism.
  \emph{Catal. Today} \textbf{2015}, \emph{244}, 115--121, Heterogeneous
  Catalysis and Surface Science\relax
\mciteBstWouldAddEndPuncttrue
\mciteSetBstMidEndSepPunct{\mcitedefaultmidpunct}
{\mcitedefaultendpunct}{\mcitedefaultseppunct}\relax
\EndOfBibitem
\bibitem[Quintas-S\'anchez \latin{et~al.}(2012)Quintas-S\'anchez, Larr\'egaray,
  Crespos, Martin-Gondre, Rubayo-Soneira, and Rayez]{Quintas2012JCP}
Quintas-S\'anchez,~E.; Larr\'egaray,~P.; Crespos,~C.; Martin-Gondre,~L.;
  Rubayo-Soneira,~J.; Rayez,~J.~C. Dynamical Reaction Pathways in Eley-Rideal
  Recombination of Nitrogen from W(100). \emph{J. Chem. Phys.} \textbf{2012},
  \emph{137}, 064709\relax
\mciteBstWouldAddEndPuncttrue
\mciteSetBstMidEndSepPunct{\mcitedefaultmidpunct}
{\mcitedefaultendpunct}{\mcitedefaultseppunct}\relax
\EndOfBibitem
\bibitem[P\'etuya \latin{et~al.}(2014)P\'etuya, Crespos, Quintas-S\'anchez, and
  Larr\'egaray]{Petuya2014JPCC}
P\'etuya,~R.; Crespos,~C.; Quintas-S\'anchez,~E.; Larr\'egaray,~P. Comparative
  Theoretical Study of H$_2$ Eley-Rideal Recombination Dynamics on W(100) and
  W(110). \emph{J. Phys. Chem. C} \textbf{2014}, \emph{118}, 11704--11710\relax
\mciteBstWouldAddEndPuncttrue
\mciteSetBstMidEndSepPunct{\mcitedefaultmidpunct}
{\mcitedefaultendpunct}{\mcitedefaultseppunct}\relax
\EndOfBibitem
\bibitem[Galparsoro \latin{et~al.}(2017)Galparsoro, Juaristi, Crespos, Alducin,
  and Larrégaray]{Galparsoro2017}
Galparsoro,~O.; Juaristi,~J.~I.; Crespos,~C.; Alducin,~M.; Larrégaray,~P.
  Stereodynamics of Diatom Formation through Eley–Rideal Abstraction.
  \emph{J. Phys. Chem. C} \textbf{2017}, \emph{121}, 19849--19858\relax
\mciteBstWouldAddEndPuncttrue
\mciteSetBstMidEndSepPunct{\mcitedefaultmidpunct}
{\mcitedefaultendpunct}{\mcitedefaultseppunct}\relax
\EndOfBibitem
\bibitem[Wu \latin{et~al.}(2019)Wu, Zhou, and Guo]{Wu2019}
Wu,~Q.; Zhou,~L.; Guo,~H. Steric Effects in CO Oxidation on Pt(111) by
  Impinging Oxygen Atoms Lead to an Exclusive Hot Atom Mechanism. \emph{J.
  Phys. Chem. C} \textbf{2019}, \emph{123}, 10509--10516\relax
\mciteBstWouldAddEndPuncttrue
\mciteSetBstMidEndSepPunct{\mcitedefaultmidpunct}
{\mcitedefaultendpunct}{\mcitedefaultseppunct}\relax
\EndOfBibitem
\bibitem[Galparsoro \latin{et~al.}(2016)Galparsoro, Pétuya, Busnengo,
  Juaristi, Crespos, Alducin, and Larregaray]{Galparsoro2016}
Galparsoro,~O.; Pétuya,~R.; Busnengo,~F.; Juaristi,~J.~I.; Crespos,~C.;
  Alducin,~M.; Larregaray,~P. Hydrogen abstraction from metal surfaces: when
  electron–hole pair excitations strongly affect hot-atom recombination.
  \emph{Phys. Chem. Chem. Phys.} \textbf{2016}, \emph{18}, 31378--31383\relax
\mciteBstWouldAddEndPuncttrue
\mciteSetBstMidEndSepPunct{\mcitedefaultmidpunct}
{\mcitedefaultendpunct}{\mcitedefaultseppunct}\relax
\EndOfBibitem
\bibitem[Galparsoro \latin{et~al.}(2017)Galparsoro, Busnengo, Juaristi,
  Crespos, Alducin, and Larregaray]{GalparsoroJCPcom2017}
Galparsoro,~O.; Busnengo,~H.~F.; Juaristi,~J.~I.; Crespos,~C.; Alducin,~M.;
  Larregaray,~P. {Communication: Hot-atom abstraction dynamics of hydrogen from
  tungsten surfaces: The role of surface structure}. \emph{J. Chem. Phys.}
  \textbf{2017}, \emph{147}, 121103\relax
\mciteBstWouldAddEndPuncttrue
\mciteSetBstMidEndSepPunct{\mcitedefaultmidpunct}
{\mcitedefaultendpunct}{\mcitedefaultseppunct}\relax
\EndOfBibitem
\bibitem[Galparsoro \latin{et~al.}(2018)Galparsoro, Busnengo, Martinez,
  Juaristi, Alducin, and Larregaray]{Galparsoro2018}
Galparsoro,~O.; Busnengo,~H.~F.; Martinez,~A.~E.; Juaristi,~J.~I.; Alducin,~M.;
  Larregaray,~P. Energy dissipation to tungsten surfaces upon hot-atom and
  Eley–Rideal recombination of H$_2$. \emph{Phys. Chem. Chem. Phys.}
  \textbf{2018}, \emph{20}, 21334--21344\relax
\mciteBstWouldAddEndPuncttrue
\mciteSetBstMidEndSepPunct{\mcitedefaultmidpunct}
{\mcitedefaultendpunct}{\mcitedefaultseppunct}\relax
\EndOfBibitem
\end{mcitethebibliography}

\end{document}